\documentclass[twocolumn]{aastex7}  

\usepackage{xcolor, amsmath}
\usepackage{comment}
\usepackage{graphicx}
\usepackage[font=small,labelfont=bf]{caption,subcaption}
\usepackage{sidecap}
\sidecaptionvpos{figure}{t}

\newcommand{\target}{WR~31a}
\newcommand{\kms}{km~s$^{-1}$}

\newcommand{\mdotsun}{\mathrm{M}_{\odot}~\mathrm{yr}^{-1}}

\shorttitle{Variable Polarization of WR~31a}
\shortauthors{Erba et al.}

\newcommand{\stsci}{Space Telescope Science Institute, 3700 San Martin Drive, Baltimore, MD 21218, USA}
\newcommand{\etsu}{Department of Physics and Astronomy, East Tennessee State University, Johnson City, TN 37663, USA}

\begin{document}

\title{Variable Polarization of WR~31a:  Binary Companion or Co-Rotating Interaction Region?}

\correspondingauthor{Christiana Erba}

\author[0000-0003-1299-8878]{Christiana Erba}
\affiliation{\stsci}
\affiliation{\etsu}
\email[show]{christi.erba@gmail.com}
\author[0000-0002-7204-5502]{Richard Ignace}
\affiliation{\etsu}
\email{ignace@etsu.edu}
\author{Faith Simmons}
\affiliation{\etsu}
\email{simmonsf1@etsu.edu}
\author{Ben Davies}
\affiliation{Independent researcher}
\email{ben.j.davies@icloud.com}

\received{2025 March 13}
\revised{2025 April 25}
\accepted{2025 April 30}


\begin{abstract}
WR~31a (Hen 3-519) is likely a post-luminous blue variable (LBV) star that is evolving to become a classical Wolf-Rayet star. Multicolor (UBVR) photopolarimetric observations of \target{} were obtained over nine nights in early 2007. The linear polarization data of \target{} trace a ``loop'' structure in a Stokes Q-U diagram, which is similar in all four passbands. After mean subtraction, the four loops align to form a single overall pattern. Such loops can be expected to arise from binary systems.
We test the binary hypothesis with two models. The data are fit for a strictly circular orbit to derive an orbital period of 16.7~d, requiring a high inclination perspective of $i\sim 80^\circ$. We also consider an elliptical orbit under simplifying assumptions, yielding a match for $i\sim 75^\circ$ with eccentricity $e\sim 0.5$ and a longer orbital period of about 70~d. The prevalence of binarity among massive stars is well-known; the prospect of detecting a binary companion during the post-LBV stage of \target{} would add to an emerging narrative of diverse interactions between massive multiple components as a function of evolutionary stage. However, if the loop originates because of a co-rotating interaction region (CIR), then the rotation period could be 8.5~d or 17~d. This would give an estimated equatorial rotation speed of 95 or 190~km/s.  Either of these is a significant fraction of the estimated critical speed of rotational break-up at 320~km/s (for an Eddington factor of $\Gamma=0$).
\end{abstract} 



\section{Introduction\label{sec:intro}}

Massive stars remain a stubbornly challenging class to decipher. The complication is twofold: massive stars are almost always born into multiple systems, and many stellar pairs are likely to interact \citep[e.g.,][]{Sana2012,Sana2014,Moe2017}. 
Direct and indirect evidence for these interactions is fairly numerous. Indirect evidence comes from massive stars that appear to be single but are ``runaways,'' suggestive of having become unbound because a companion exploded as a supernova \citep[e.g.,][]{2019A&A...624A..66R,2022MNRAS.516.2142A}. Direct evidence comes from systems with active mass transfer, as seen in $\beta$ Lyr \citep{2002AN....323...87H} or indications of past interactions as in the relatively common population of Be~stars \citep{2024ApJ...962...70K,2025A&A...694A.208K}. The latter example consists of rapidly rotating massive stars, typically on the main sequence or red giant branch, that have Keplerian disks \citep{2013A&ARv..21...69R}. There is now growing evidence of stripped core companions to the rapidly rotating B primaries, which are the remains from a previous episode of mass transfer \citep{2019ApJ...885..147K}. 
Ultimately, these interactions are consequential to local stellar and galactic evolution: the high luminosities, powerful stellar winds, and explosive terminations of massive stars substantially contribute to the chemical and dynamical enrichment of their locales \citep[e.g.,][and references therein]{Livio_Villaver_2009,2012MNRAS.421.3522H}, and leave behind exotic compact remnants such as neutron stars and black holes \citep{2019A&A...621A..92S,2021MNRAS.505.4874H}.  

An even greater challenge is presented by the particularly rare subgroup of evolved massive stars known as luminous blue variables (LBVs). The LBV phase is an intermediate, short-lived ($\sim$25~kyr) stage of massive O star evolution that presumably precedes the Wolf-Rayet (WR) stage \citep[e.g.,][and sources therein]{1994PASP..106.1025H,1996LIACo..33...39M}. 
During quiescence, LBVs are hot and extremely luminous\footnote{Quiescent LBVs along the S Dor instability strip can average 20-40~kK in temperature; however, during eruptive events, LBVs are much cooler at around 8-10 kK \citep[e.g.,][]{Smith2017_rev}.}, with the star often situated near the Humphreys-Davidson Limit for stellar luminosity. LBVs are distinguished by their observed photometric variability and their significant, irregular eruptive outbursts that shed the outer layers of the stellar photosphere \citep{1994PASP..106.1025H, 2011MNRAS.415..773S}. Such eruptions are typically categorized by timescale or size: for example, as ``semiregular'' eruptions \citep[e.g., S~Dor][]{2001A&A...366..508V}, or ``giant'' eruptions \citep[e.g., $\eta$ Car;][]{2020Galax...8...20W}. Even during the quiescent phase in between eruptive events, LBVs rapidly shed mass at rates of around $10^{-5}~\mdotsun$ \citep{1994PASP..106.1025H}. Such extreme mass loss rates fundamentally determine the evolution of the LBV and its immediate environment. In particular, recent work has shown that slower, dense massive star winds, such as those found in LBV and WR stars, play a primary role in interstellar dust production \citep{Agliozzo2021} and in the chemical enrichment of globular clusters \citep{Vink2018}. 

The subject of this study is \target{} (Hen~3-519, WRAY~15-682), a candidate LBV star that displays a Wolf-Rayet (WR) emission line spectrum. \citet{Smith94} noted several similarities between the spectra of \target{} and AG~Car, classifying both as spectral type WN11. \target{} is surrounded by a ring nebula, 1' in diameter \citep{Smith94}, that has a characteristic ``WR bubble'' shape \citep{2015A&A...578A..66T}. The nebula presumably provides evidence of a significant eruptive event earlier in the star's evolutionary history. To date, a binary companion has not been identified for \target, and the star is considered to be well-isolated from any other massive stars \citep{2015MNRAS.447..598S,2024A&A...692A.109D}.   

\citet{Davidson93} proposed that \target{} is transitioning between an LBV and WN evolutionary stage.
While LBVs are generally accepted to be evolutionary precursors to WRs, the details of these proposed pathway(s) are still unclear. Recent multiplicity surveys of WRs, such as those by \citet{2020A&A...641A..26D,2022A&A...664A..93D,2023A&A...674A..88D} and \citet{2024A&A...692A.109D}, have estimated the binary fraction for WN stars to be at least 30\% and possibly up to 50\%, with higher rates of binarity among the WC star population. A similar study by \citet{2022A&A...657A...4M} found the binary fraction among galactic LBV stars to be just below the 30\% threshold.
However, these rates also reveal that multiplicity is not uniformly observed across evolved massive stars, and indeed, (presumably) single LBVs and WRs like \target{} have been identified. For these systems, two questions remain: is a binary companion is present, but as yet unobserved; or did a binary companion preexist the currently observed state? Studies from \citet{2015MNRAS.447..598S}, \citet{2017MNRAS.472..591A}, and \citet{2022MNRAS.516.2142A} all agree that the binary scenario is the most consistent interpretation of the evolutionary history for LBVs (and thus WR stars). Within this context, the possibility of observing a binary companion around a potentially transitionary object like \target{}, and to determine its orbital and companion properties, is certainly paramount.

Linear polarimetry is a powerful tool for inferring geometry in sources that are spatially unresolved \citep{2010stpo.book.....C}. This is because sources with centro-symmetric intensity distributions produce no net polarization; consequently, any net polarization implies a deviation from spherical symmetry. Variable polarization arises from binary scenarios as a function of orbital phase. LBV and WR stars have been popular targets for polarization studies due to their evolved evolutionary stage, the strength of their winds, and the occurrence of binarity. In particular, polarimetry has been used to better understand the connection between bipolar nebulae and the wind driving or presence of binarity in LBVs \citep[e.g.][]{1993ApJ...407..723S, 1994ApJ...429..846S, 1995AJ....110..251C}. The clumpy nature of LBV winds has also been studied through polarimetric fluctuations \cite[e.g., ][]{2005A&A...439.1107D, 2006PASP..118..820W, 2020ApJ...900..162G}. Similar studies exist for the WR stars with colliding wind binaries \citep[e.g.,][]{1999PASP..111..861E, 2022ApJ...933....5I}, component masses in binaries \citep[e.g.,][]{2022ApJ...930...89F}, aspherical wind geometry \citep[e.g.,][]{2005ApJ...623.1092V}, and wind clumping \citep[e.g.,][]{2023MNRAS.519.3271I}.

We here report on new polarimetric observations of \target. The pattern described by the polarization data is suggestive of a binary companion or a co-rotating interaction region (CIR). We explore the prospect of a binary companion. Section~\ref{sec:data} reports the observations. In section~\ref{sec:binary}, we present two models for deriving orbital parameters for the proposed companion from the polarimetric measurements. Concluding remarks are discussed in section~\ref{sec:concl}.


\section{Polarimetry \label{sec:data}}

\begin{deluxetable*}{lccccccccr}
\tablecaption{Journal of observations and polarization data for \target. The Bessell U filter (U\#640) corresponds to a wavelength range of 3102.09 - 3980.28 \AA. The Bessell B filter (B\#639) corresponds to a wavelength range of 3592.42 - 5738.27 \AA. The O\textsc{III} filter (O\textsc{III}\#687) corresponds to a wavelength range of 4941.24 - 5074.24 \AA. The H$\alpha$ Red filter (HalR\#709) corresponds to a wavelength range of 6568.52 - 6718.30 \AA.    
\label{tab:observations}}
\tablehead{
\colhead{MJD} & \colhead{Filter} & \colhead{Exp} & \colhead{$p$} & \colhead{$\psi_p$} & \colhead{$q$} & \colhead{$\Delta q$} & \colhead{$u$} & \colhead{$\Delta u$} & \colhead{$\sigma_{q,u}$} \\
\colhead{(JD-2400000.5)} & \colhead{} & \colhead{(s)} & \colhead{(\%)} & \colhead{(${^\circ}$)} & \colhead{(\%)} & \colhead{(\%)} & \colhead{(\%)} & \colhead{(\%)} & \colhead{(\%)}
}
\startdata
54151.2980 & U\#640 & 40 & 2.324 & 99.390  & $-$2.200 & $-$2.112 & $-$0.748 & $-$0.698 & 0.039 \\
54153.2123 & U\#640 & 20 & 2.399 & 96.260  & $-$2.342 & $-$2.112 & $-$0.520 & $-$0.698 & 0.070 \\
54154.1957 & U\#640 & 20 & 2.416 & 95.560  & $-$2.370 & $-$2.112 & $-$0.466 & $-$0.698 & 0.092 \\
54155.2455 & U\#640 & 20 & 2.259 & 96.840  & $-$2.195 & $-$2.112 & $-$0.534 & $-$0.698 & 0.072 \\
54156.2369 & U\#640 & 20 & 2.367 & 100.000 & $-$2.224 & $-$2.112 & $-$0.810 & $-$0.698 & 0.048 \\
54157.2211 & U\#640 & 20 & 2.267 & 101.690 & $-$2.081 & $-$2.112 & $-$0.900 & $-$0.698 & 0.057 \\
54158.2177 & U\#640 & 20 & 2.187 & 102.930 & $-$1.968 & $-$2.112 & $-$0.954 & $-$0.698 & 0.054 \\
54159.2257 & U\#640 & 20 & 1.887 & 102.080 & $-$1.721 & $-$2.112 & $-$0.772 & $-$0.698 & 0.057 \\
54160.2266 & U\#640 & 20 & 1.995 & 98.390  & $-$1.910 & $-$2.112 & $-$0.576 & $-$0.698 & 0.047 \\
54151.3069 & B\#639 & 20 & 2.756 & 98.490  & $-$2.636 & $-$2.596 & $-$0.805 & $-$0.799 & 0.030 \\
54153.2191 & B\#639 & 10 & 2.733 & 96.490  & $-$2.663 & $-$2.596 & $-$0.614 & $-$0.799 & 0.026 \\
54154.2024 & B\#639 & 10 & 2.798 & 95.650  & $-$2.744 & $-$2.596 & $-$0.548 & $-$0.799 & 0.029 \\
54155.2548 & B\#639 & 10 & 2.843 & 97.370  & $-$2.749 & $-$2.596 & $-$0.723 & $-$0.799 & 0.039 \\
54156.2462 & B\#639 & 10 & 2.902 & 99.520  & $-$2.743 & $-$2.596 & $-$0.947 & $-$0.799 & 0.032 \\
54157.2301 & B\#639 & 10 & 2.767 & 99.940  & $-$2.602 & $-$2.596 & $-$0.941 & $-$0.799 & 0.037 \\
54158.2272 & B\#639 & 7 & 2.575 & 101.220 & $-$2.380 & $-$2.596 & $-$0.983 & $-$0.799 & 0.030 \\
54159.2344 & B\#639 & 7 & 2.425 & 100.490 & $-$2.264 & $-$2.596 & $-$0.868 & $-$0.799 & 0.032 \\
54160.2357 & B\#639 & 7 & 2.692 & 98.250  & $-$2.582 & $-$2.596 & $-$0.765 & $-$0.799 & 0.024 \\
54151.3145 & O\textsc{III}\#687 & 30 & 2.875 & 98.110  & $-$2.761 & $-$2.758 & $-$0.803 & $-$0.809 & 0.050 \\
54153.2251 & O\textsc{III}\#687 & 15 & 2.905 & 96.620  & $-$2.827 & $-$2.758 & $-$0.665 & $-$0.809 & 0.072 \\
54154.2084 & O\textsc{III}\#687 & 15 & 2.986 & 95.490  & $-$2.931 & $-$2.758 & $-$0.569 & $-$0.809 & 0.055 \\
54155.2633 & O\textsc{III}\#687 & 15 & 2.954 & 96.780  & $-$2.872 & $-$2.758 & $-$0.693 & $-$0.809 & 0.059 \\
54156.2546 & O\textsc{III}\#687 & 15 & 3.027 & 98.790  & $-$2.886 & $-$2.758 & $-$0.914 & $-$0.809 & 0.057 \\
54157.2385 & O\textsc{III}\#687 & 18 & 2.936 & 99.270  & $-$2.784 & $-$2.758 & $-$0.934 & $-$0.809 & 0.055 \\
54158.2355 & O\textsc{III}\#687 & 18 & 2.858 & 100.440 & $-$2.671 & $-$2.758 & $-$1.019 & $-$0.809 & 0.046 \\
54159.2427 & O\textsc{III}\#687 & 18 & 2.652 & 100.090 & $-$2.489 & $-$2.758 & $-$0.915 & $-$0.809 & 0.057 \\
54160.2441 & O\textsc{III}\#687 & 18 & 2.716 & 98.230  & $-$2.605 & $-$2.758 & $-$0.770 & $-$0.809 & 0.052 \\
54151.3230 & HalR\#709 & 30 & 3.136 & 97.470  & $-$3.030 & $-$2.939 & $-$0.808 & $-$0.804 & 0.030 \\
54153.2316 & HalR\#709 & 15 & 3.074 & 96.280  & $-$3.000 & $-$2.939 & $-$0.668 & $-$0.804 & 0.030 \\
54154.2148 & HalR\#709 & 15 & 2.974 & 95.420  & $-$2.921 & $-$2.939 & $-$0.559 & $-$0.804 & 0.037 \\
54155.2723 & HalR\#709 & 15 & 3.046 & 96.340  & $-$2.971 & $-$2.939 & $-$0.669 & $-$0.804 & 0.031 \\
54156.2637 & HalR\#709 & 15 & 3.213 & 98.150  & $-$3.084 & $-$2.939 & $-$0.902 & $-$0.804 & 0.028 \\
54157.2476 & HalR\#709 & 12 & 3.146 & 99.120  & $-$2.988 & $-$2.939 & $-$0.985 & $-$0.804 & 0.033 \\
54158.2446 & HalR\#709 & 12 & 3.047 & 99.630  & $-$2.876 & $-$2.939 & $-$1.005 & $-$0.804 & 0.046 \\
54159.2520 & HalR\#709 & 12 & 2.857 & 98.560  & $-$2.730 & $-$2.939 & $-$0.841 & $-$0.804 & 0.033 \\
54160.2532 & HalR\#709 & 12 & 2.963 & 97.840  & $-$2.853 & $-$2.939 & $-$0.801 & $-$0.804 & 0.040 \\
\enddata
\tablecomments{Additional details on the filters are available from the SVO Filter Profile Service \citep{SVO-1,SVO-2,SVO-3}.}
\end{deluxetable*}

Photopolarimetry of \target{} was obtained over a span of nine nights between 2007~February~20 - March~1 with the ESO Faint Object Spectrograph and Camera (EFOSC2), mounted on the ESO 3.6~m Telescope at La Silla Observatory in Chile. The instrumental configuration applied the "IN" position setting for the half-wave plate and utilized the 20'' Wollaston Prism, which separates the beam into orthogonal {\it O} and {\it E} components. The waveplate is rotated through four different angles (0$^\circ$, 22.5$^\circ$, 45$^\circ$, 67.5$^\circ$) to obtain the normalized Stokes parameters q and u \citep[e.g.,][]{2010stpo.book.....C}. The degree of polarization ($p$) and position angle (PA) are measured from the normalized Stokes parameters using
\begin{eqnarray}
\label{eq:pol}
p & = & \sqrt{q^2+u^2},\\
\psi_p & = & \frac{1}{2} \arctan\left(\frac{u}{q}\right).
\end{eqnarray}
\target{} was observed with a set of four narrowband filters, with exposure times adjusted per night based on the weather conditions. The journal of observations and the filter details are reported in Table~\ref{tab:observations}.

\begin{figure*}[t]
    \centering
    \begin{subfigure}[b]{0.48\textwidth}
    \centering
    \includegraphics[width=\columnwidth]{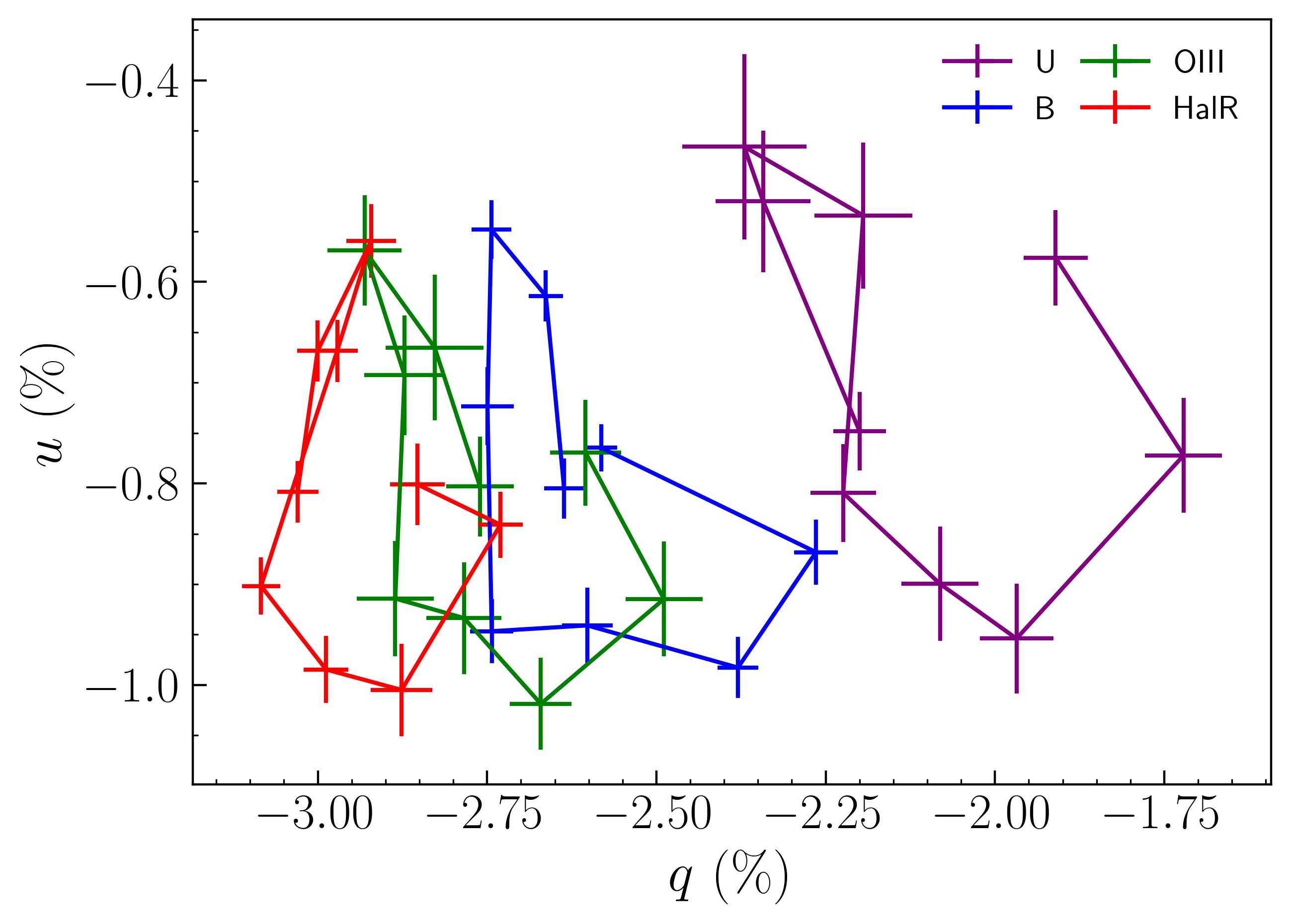}
    \end{subfigure}
     \hfill
    \begin{subfigure}[b]{0.48\textwidth}
    \centering
    \includegraphics[width=\columnwidth]{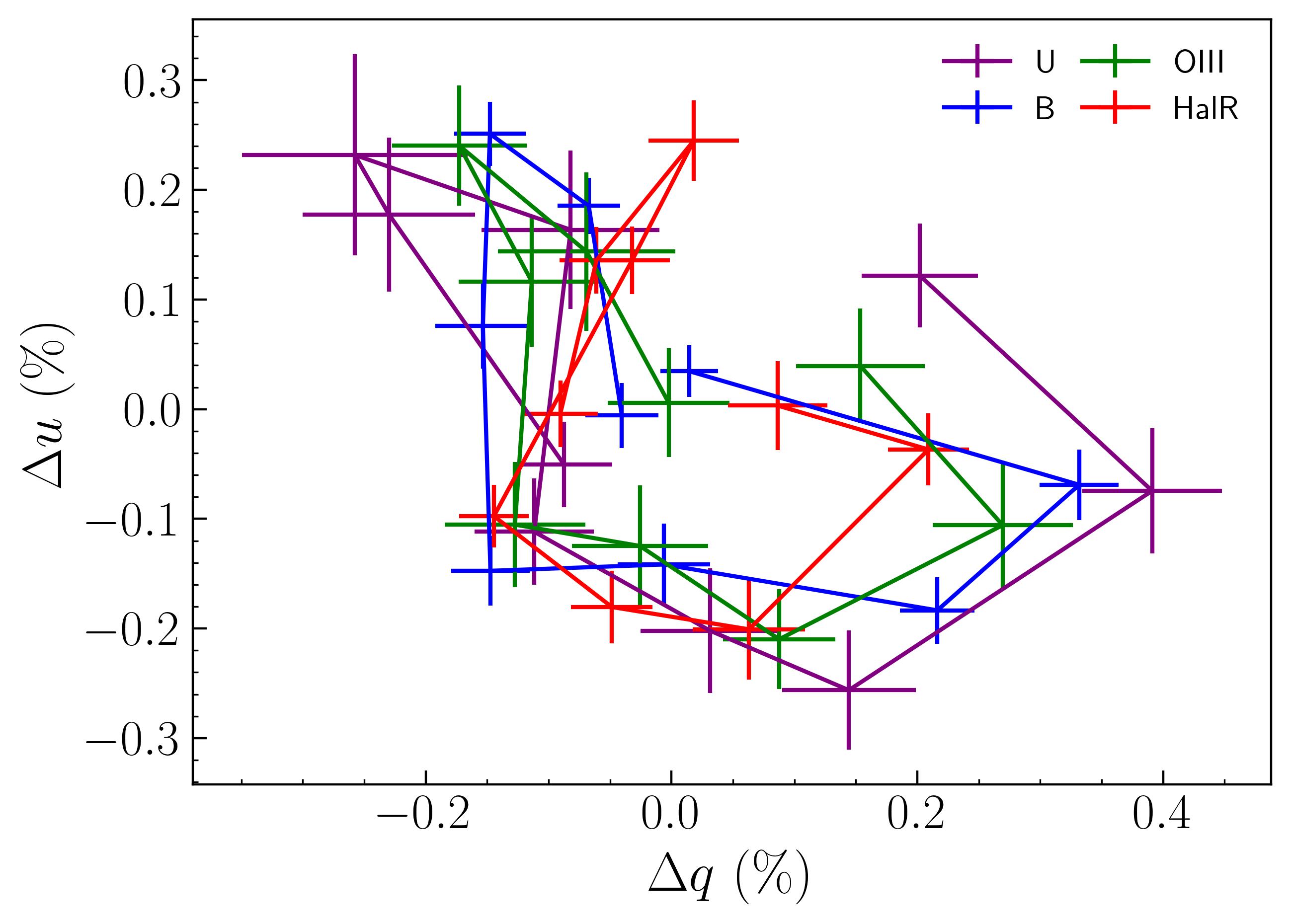}
    \end{subfigure}
    \caption{$q-u$ diagrams illustrating the time-variable polarization of \target, with filters separated by color (see Table~\ref{tab:observations}). The left panel shows the diagram before the means (eq.~\ref{eq:meanqu}) are subtracted; the right panel shows the diagram post-mean subtraction (eq.~\ref{eq:dqu}).}
    \label{fig:qu_plots}
\end{figure*}

Figure~\ref{fig:qu_plots} (left) shows the UBVR\footnote{Here and throughout, ``V'' corresponds to the O\textsc{III}\#687 filter and ``R'' corresponds to the HalR\#709 filter. See Table~\ref{tab:observations} for more details.} photopolarimetry in a $q-u$ diagram. Each of the passbands notably displays the same characteristic loop morphology, with essentially the same form and amplitude. The right panel shows the same diagram, but with the means of each passband subtracted, producing the shifted values $\Delta q$ and $\Delta u$. We introduce subscript ``i'' for the passbands UBVR, and ``j'' for the measurements 1--9.
The mean of each passband is calculated using 
\begin{eqnarray}
\label{eq:meanqu}
\bar{q_{\rm i}} & = \frac{1}{9}\,\sum_1^{9} q_{\rm i,j} \\
\bar{u_{\rm i}} & = \frac{1}{9}\,\sum_1^{9} u_{\rm i,j}.
\end{eqnarray}
\noindent The shifted polarizations are then
\begin{eqnarray}
\label{eq:dqu}
\Delta q_{\rm i,j} & = q_{\rm i,j} - \bar{q_{\rm i}} \\    
\Delta u_{\rm i,j} & = u_{\rm i,j} - \bar{u_{\rm i}}.
\label{eq:dquu}
\end{eqnarray}
The interstellar polarization (ISP) $p_I$ and position angle $\psi_I$, or equivalently the Stokes parameters $q_I$ and $u_I$, are not known. However, the ISP is not time-dependent, so the time variability of the polarization shown in the right panel of Figure~\ref{fig:qu_plots} is certainly stellar in origin. The mean subtraction thus approximates a correction for the ISP. 

The alignment of the loops in each passband is notable, but is generally expected as the result of electron scattering, which produces a gray opacity at visible wavelengths \citep{1978A&A....68..415B}. This is the main argument in favor of binarity: in the presence of pure electron scattering, a binary system is expected to produce a loop morphology with a shape and amplitude that is independent of wavelength. This is in fact what is seen for \target.


\section{Constraining Orbital Parameters from Best-Fit Models \label{sec:binary}}

The polarimetric data for \target{} are highly suggestive of binarity. This is a falsifiable claim: if a binary companion were present, the loop morphology should repeat. It will thus be important to obtain new observations as a direct test of this claim. For the interim, we assume binarity in order to examine what the existing data imply about the properties of a companion and its orbit. 

With a constrained dataset and a loop that is {\it nearly} complete, we can imagine two limiting test cases: a circular orbit and an elliptical orbit. For optically thin electron scattering, the former has been discussed in detail by \citet{1978A&A....68..415B} and the latter has been approached in a special case by \citet{1982MNRAS.198..787B}. This second approach is quite general, allowing for the treatment of single scattering in arbitrary envelopes with any number of point-like sources of illumination; thus, we use this formalism to handle both of our test cases. 

\begin{deluxetable}{llr}
\tablecaption{Properties of \target \label{tab:hen3-519_properties}}
\tablehead{
\colhead{Property} & \colhead{Value} & \colhead{Source}
}
\startdata
$\log L_\ast/L_\odot$ & 5.8 & \citet{2019MNRAS.488.1760S} \\
$T_{\rm eff}^a$ & 27,500 K & \citet{Smith94}\\
$\dot{M}$ & $12\times 10^{-5}~M_\odot$/yr& \citet{Smith94}\\
$M_\ast$ & $17~M_\odot$ & \citet{Smith94}\\
$R_\ast^a$ & $31.6~R_\odot$ & \citet{Smith94}\\
$v_\infty$ & 365 km/s & \citet{Smith94}\\ 
dist. & $7.6^{+2.5}_{-1.7}$ kpc
  & \citet{2019MNRAS.488.1760S}\\ 
\enddata
\tablecomments{$^a$As a candidate LBV, temperature and radius are subject to variations. Quoted values are adopted as reasonably applying to the time when the polarimetric data were obtained.}
\end{deluxetable}

We acknowledge that such calculations fail to capture the effects of clumping, which for a (time-averaged) spherically symmetric wind will produce a scatter plot in a $q-u$ diagram \citep[c.f. Fig.~1 of][]{2025arXiv250402659I}. The combination of binarity and wind clumping has not been emphasized in previous modeling studies. There are scenarios where either could dominate the signal. Given that we observe what appears to be a $q-u$ loop for \target{} that is the same for different wavelengths, we assume as a working hypothesis that there is a binary signal which may be influenced (but not dominated by) clumping. The impact of clumping might then serve to distort the loop, which could affect inferences of system inclination and/or orbital eccentricity. This could mean that with multiple orbits, no two loops would be identical, and yet their average in the $q-u$ plane may more precisely define the orbit. We remind the reader of this possibility in our concluding remarks, but proceed with an analysis that ignores wind clumping effects.

\citet{1982MNRAS.198..787B} derive expressions for a binary system with an elliptical orbit of eccentricity $e$. The circular orbit solution is thus straightforwardly derived by setting $e=0$. The model makes additional simplifying assumptions beyond that of the earlier \citet{1978A&A....68..415B} work, specifically that (a) the primary star is the dominant source of illumination for scattered light; and (b) symmetry breaking arises from scatterers close to the secondary star. The first assumption implies either a small, low-mass companion, or that the companion is faint compared to the primary at the wavelengths where the polarimetric data are obtained. The second assumption essentially localizes the production of net polarization to track with the orbit of the secondary.

When two massive stars are in a binary, it is common for them to have a colliding wind interaction \citep[e.g.,][]{1992ApJ...386..265S} in which the wind of one star largely dominates that of the other. 
\target{} is a luminous star with a high mass-loss rate (see Table~\ref{tab:hen3-519_properties}); whatever the purported companion, its wind is likely to be dominated by the primary. Consequently, if the wind of \target{} is basically spherical (at least in time average), it can be reasonably assumed that the breaking of that symmetry is localized to the location of the companion.

It is exactly this situation for which \cite{1982MNRAS.198..787B} derived their analytic formula. Assuming that symmetry breaking is near the secondary star, the analytic solution for the orbital-phase-dependent variation of polarization involves the first through fourth harmonics, whereas the circular orbit only has the second harmonic. The differential relative Stokes polarizations $\Delta q$ and $\Delta u$ are given by
\begin{eqnarray}
\Delta q & = & \frac{-\tau_\ast}{(1-e^2)^2}\,(1+\cos^2 i)\,
    \left\{ \frac{e^2}{4}\,\cos 2\phi_{\rm p} + e\,\cos(\phi+\phi_{\rm p}) \right. \nonumber \\
    & & \left. +\left(1+\frac{e^2}{2}\right) \,\cos (2\phi) +3e\,\cos(3\phi-\phi_{\rm p}) \right. \nonumber \\
    & & \left. + \frac{e^2}{4}\,\cos[2(2\phi-\phi_{\rm p})] \right\} \label{eq:dq}  \\
\Delta u & = & \frac{-\tau_\ast}{(1-e^2)^2}\,(2\cos i)\,
    \left\{ \frac{e^2}{4}\,\sin 2\phi_{\rm p} + e\,\sin(\phi+\phi_{\rm p}) \right. \nonumber \\
    & & \left. +\left(1+\frac{e^2}{2}\right) \,\sin (2\phi) +3e\,\sin(3\phi-\phi_{\rm p}) \right. \nonumber \\
    & & \left. + \frac{e^2}{4}\,\sin[2(2\phi-\phi_{\rm p})] \right\} . \label{eq:du}
\end{eqnarray}
\noindent The various factors in these expressions are $\phi$ for the longitude of the scattering region (i.e., the secondary star, in our case) with respect to the Earth, $\phi_{\rm p}$ for the longitude of periastron passage, $\tau_\ast$ for the optical depth for the scattering region, and $i$ for the viewing inclination. The orbit of the secondary about the primary is given by
\begin{equation}
r(\phi) = a \left[\frac{1-e^2}{1+e\,\cos(\phi-\phi_{\rm p})}\right],
\end{equation}
\noindent where $a$ is the semi-major axis.  These equations and parameter definitions can be found in \cite{1982MNRAS.198..787B}.

Before further exploring applications of circular and elliptical orbits to explain the observations, some general comments are in order. First, the expression $\tau_\ast/(1-e^2)^2$ is a scaling factor common to both $\Delta q$ and $\Delta u$.  Second, the expression for $\Delta q$ consists of cosine terms up to the fourth harmonic, and similarly the expression for $\Delta u$ consists of sine terms to the fourth harmonic. Consequently, matching harmonic terms between the two, each would produce a circle as seen from a pole-on perspective. In combination, these circular contributions are scaled by respective functions of eccentricity. Third, except when the eccentricity approaches unity, the terms that scale with $e^2$ will have minor contributions to the polarimetric variability. Finally, the expressions in eqs.~(\ref{eq:dq}) and (\ref{eq:du}) lead to repeating loop-like variations in a traditional $Q-U$ diagram, with inclination setting the overall ``flattening'' of the loop.  Note that $\Delta u$ goes to zero for an edge-on perspective of the orbit with $i=90^\circ$.  Consequently, the overall aspect ratio of an observed loop is set by viewing inclination, as well-studied by \cite{1978A&A....68..415B}.


\subsection{Analysis for a Circular Orbit}

\begin{figure*}[t]
    \centering
    \begin{subfigure}{0.32\textwidth}
    \includegraphics[width=\textwidth]{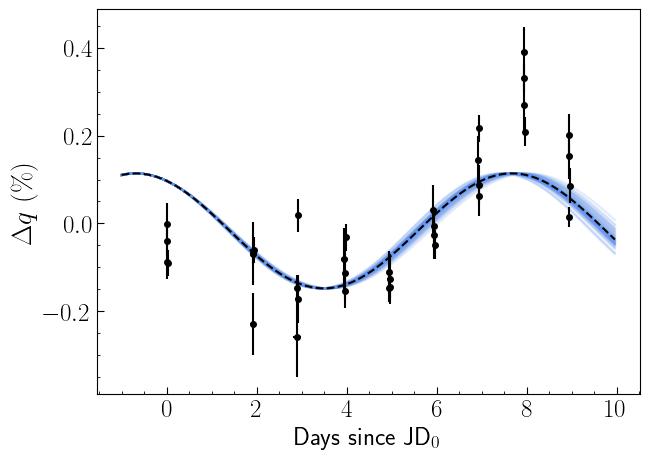}    
    \end{subfigure}
    \begin{subfigure}{0.32\textwidth}
    \includegraphics[width=\textwidth]{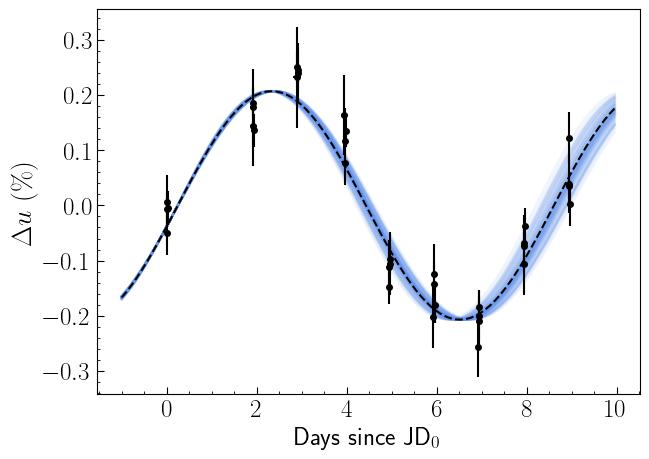}    
    \end{subfigure}
    \begin{subfigure}{0.32\textwidth}
    \includegraphics[width=\textwidth]{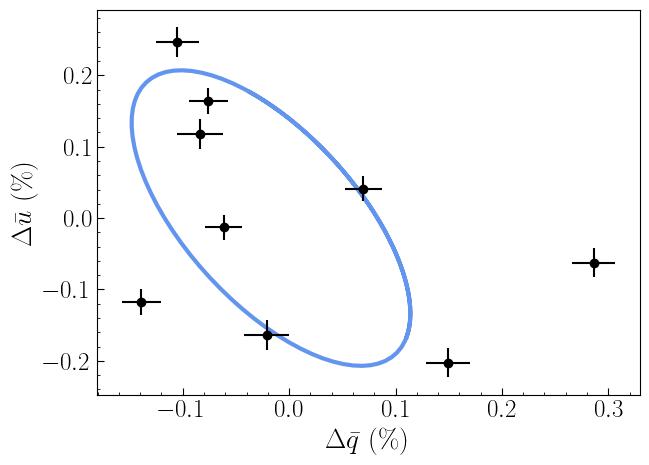}    
    \end{subfigure}
    \caption{Independent sinusoidal fits to $\Delta q$ (upper panel) and $\Delta u$ (middle panel), assuming a circular orbit. The variations in $\Delta u$ display a more prominent sinusoidal pattern, so we used its fit to set the period for the fit to $\Delta q$. The shading indicates the uncertainty in the period, fit to ${\rm P} = 16.7 \pm 0.3$~d. Time is represented in days advancing from the first day of observation (JD$_0$). The lower panel shows the resulting model in comparison to the daily weighted means of the data.
    \label{fig:qufits}}
\end{figure*}

In the case of a circular orbit, $e=0$ and only the second harmonic terms remain in eqs.~(\ref{eq:dq}) and (\ref{eq:du}). The loop is thus an ellipse with aspect ratio set by the viewing inclination as $2\cos i/ (1+\cos^2 i)$, ranging from 0~to~1.

Due to the similarity of the loop structures in the $\Delta q$-$\Delta u$ diagram of Figure~\ref{fig:qu_plots}, we combined the data from each passband to examine the variability of the shifted polarization with time. The lightcurves are shown in Figure~\ref{fig:qufits} (upper and middle panels), where time is calculated as days advancing from the first date of observation (JD$_0$), which acts as a proxy of the orbital period. The lower panel displays the resulting best-fit loop in the q-u plane, as compared with the weighted mean of the shifted polarization values, binned per night.
The fits to the polarization lightcurves (black dashed lines in Figure~\ref{fig:qufits}) were forced to be sinusoids, since that is the form predicted by \cite{1978A&A....68..415B} for binarity. More specifically, it is the expected variation when the scattering envelope around the binary is top-down symmetric about the orbital plane, and for left-right symmetry within the orbital plane with respect to the line-of-centers joining the stars. Given the limited number of measurements and the errors, these assumptions provide a reasonable match the dataset.

The best fit to the $\Delta u$ lightcurve, where the variability of the polarization appears more sinusoidal, results in a period (P) of 16.7$~\pm~$0.3~d. The blue shading around the best-fit curves in Figure~\ref{fig:qufits} indicates the uncertainty on the period, which was computed using the Markov chain Monte Carlo code \texttt{emcee} \citep{Foreman-Mackey_2013}. This reflects a formal statistical uncertainty without accounting for the possibility of distortions to the loop from wind instabilities like clumping. Note that a circular orbit would show two loops in the $q-u$ diagram for each orbit; since the data display only one loop, this represents only half of the orbital period. 

In other WN binary systems (e.g., WR~145: \citealt{Muntean2009};
WR~62a: \citealt{Collado2013};
WR~97: \citealt{Niemela1995,Niemela1996}) that have been observed\footnote{This comparison sample is small, since it is inherently constrained by the complexities of identifying both the primary and companion stars in a WR binary system (especially those of presumably analogous spectral and evolutionary type, see e.g., \citealt{Hamann2007,Shara2022}). Thus, our current understanding of such systems may be limited by the biases imposed by small-number statistics.}, the companion star in the pair is often of spectral type O5-7. For these systems, the WN star is technically the {\it secondary}, at roughly a 1:2 mass ratio with the O-type primary.
If we make a similar assumption for \target, and taking the mass of \target{} from Table~\ref{tab:hen3-519_properties}, the implied orbital radius is $r = 7.1\times 10^{12}$~cm, or about $3.2 R_\ast$. The best fit curves also lead to amplitudes in the sky system of 0.13 in $\Delta q$ and 0.21 in $\Delta u$. However, in the star system, $\Delta u_\ast \le \Delta q_\ast$ for all viewing inclinations, therefore we can assume those amplitudes in $\Delta q_\ast$ and $\Delta u_\ast$, respectively are given the following ratio:
\begin{equation}
\frac{2\cos i}{1+\cos^2 i} = \frac{0.15}{0.47}.
\end{equation}
\noindent This is a quadratic equation in $\cos i$, producing an inclination of $i=80^\circ$; however, there are significant uncertainties on this value. What seems clear is that the orbit, if circular, produces a loop in the observed $\Delta q - \Delta u$ diagram, and so is inconsistent with an edge-on inclination of $i=90^\circ$. The loop is also clearly not a circle, and so is inconsistent with a pole-on inclination of $i=0^\circ$.

\subsection{Analysis for an Elliptical Orbit}

\cite{1982MNRAS.198..787B} gives a discussion of modeling the variable polarization with an elliptical orbit as an extension to the foundational work provided by \cite{1978A&A....68..415B}. They derive an analytic solution for $q(t)$ and $u(t)$ under special circumstances. Massive stars typically have fast winds and relatively large mass-loss rates $\dot{M}$. While the majority of massive stars have a modest range of wind terminal speeds of between 1000-3000 km/s \citep[for OB stars; e.g.,][]{1990ApJ...361..607P}, there is a far more dynamic range in values of $\dot{M}$. For example, mid-spectral range B stars may have $\dot{M} < 1\times 10^{-10}~M_\odot$/yr, while O stars can rise to levels of $\dot{M} \sim 1\times 10^{-6}~M_\odot$/yr, and evolved massive stars of LBV and WR types exhibit yet greater mass-loss rates of up to $\dot{M} \sim 1\times 10^{-4}~M_\odot$/yr \citep[e.g.,][]{1994PASP..106.1025H}.

It is still the case that multiple loops in the $q-u$ diagram are expected for an elliptical orbit.  However, for an elliptical orbit, the loops deriving from periastron passage versus apastron passage can be quite different in amplitude.  For higher eccentricities, the amplitude of polarimetric variation around apastron is significantly smaller than around periastron.  As the opposite limiting case from a circular orbit, we assume that the data for \target{} is the loop from periastron passage.  Under this assumption, a longer orbital period will result, since periastron passage is short.  

An elliptical orbit also provides for more model parameters, including viewing inclination $i$, orbital eccentricity $e$, orbital period, and the longitude of periastron passage $\phi_{\rm p}$.  We carried out a modest parameter study for these variables.  Our goal is to match the overall variability of polarization with time.

\begin{figure*}[htp]
    \centering
    \begin{subfigure}[t]{0.47\textwidth}
        \centering
        \includegraphics[width=\columnwidth]{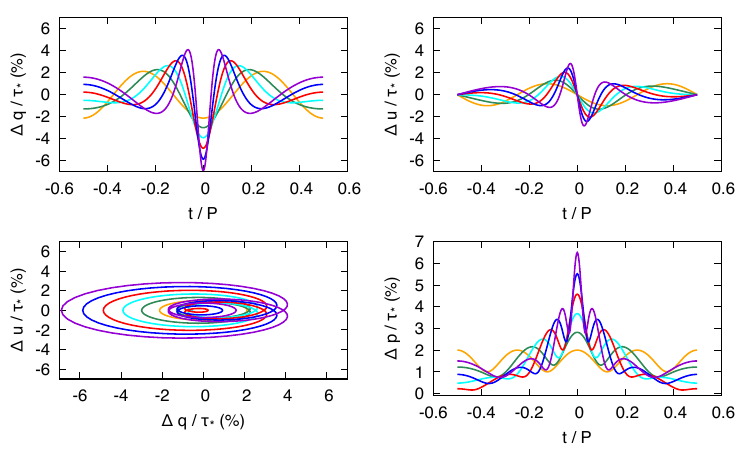} 
        \caption{Models with viewing inclination $i=75^\circ$ and the longitude of periastron as $\phi_{\rm p}=0^\circ$. Panels show (a) $\Delta q$ with orbital phase $t/P$, (b) $\Delta u$ with orbital phase, (c) $\Delta u$ with $\Delta q$, and (d) $\Delta p$ with orbital phase. Note that $\Delta p$ shows a triple peak near periastron passage at zero phase.  The polarization quantities are in percent and also normalized to $\tau_\ast$. The five curves are for eccentricities $e=$0.0, 0.1, 0.2, 0.3, 0.4, and 0.5, as orange, green, cyan, red, blue, and violet.}
        \label{fig:ecc0}
    \end{subfigure}
    \hfill
    \begin{subfigure}[t]{0.47\textwidth}
        \centering
        \includegraphics[width=\columnwidth]{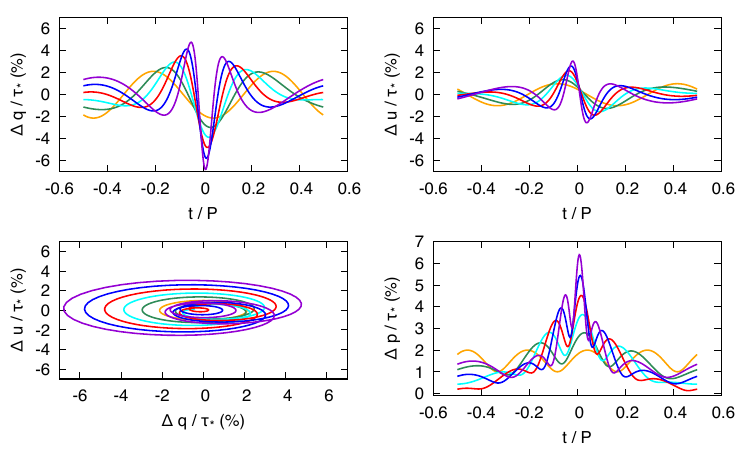}  \caption{As in Fig.~\ref{fig:ecc0}, now with $\phi_{\rm p}=130^\circ$.  Note that $\Delta p$ shows a double peak near periastron passage at zero phase, with a lower peak preceding a higher one.}
        \label{fig:ecc1}
    \end{subfigure}
    \vfill
    \begin{subfigure}[b]{0.47\textwidth}
        \centering
        \includegraphics[width=\columnwidth]{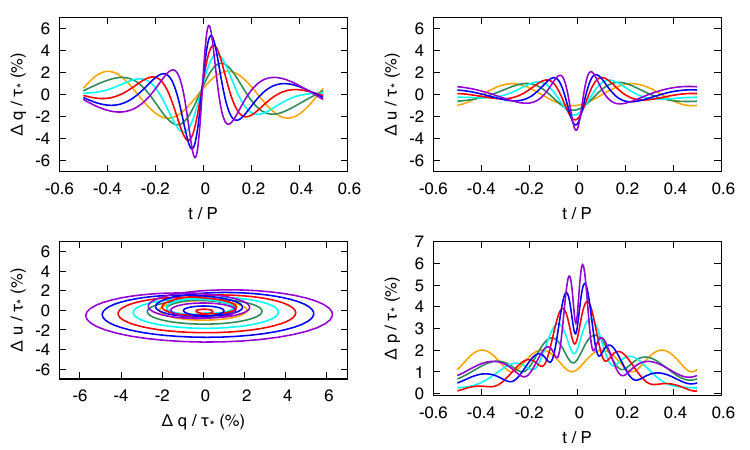} 
        \caption{As in Fig.~\ref{fig:ecc0}, now with $\phi_{\rm p}=140^\circ$. Note that $\Delta p$ shows a double peak near periastron passage at zero phase, with roughly equal peaks.}
        \label{fig:ecc2}
    \end{subfigure}
    \hfill
    \begin{subfigure}[b]{0.47\textwidth}
        \centering
        \includegraphics[width=\columnwidth]{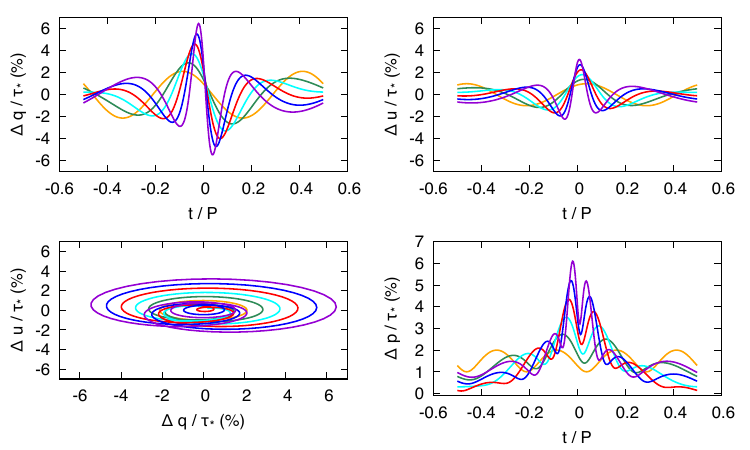} 
        \caption{As in Fig.~\ref{fig:ecc0}, now with $\phi_{\rm p}=150^\circ$. Note that $\Delta p$ shows a double peak near periastron passage at zero phase, with a higher peak preceding a lower one.}
        \label{fig:ecc3}
    \end{subfigure}
    \caption{Model calculations based on \cite{1982MNRAS.198..787B} as described in the text.  The different curves are for eccentricities $e=0.0, 0.1, 0.2, 0.3$, and 0.4.  The inclination is pole-on. Model parameters are described in the individual panel captions.}
\end{figure*}

Constraining a best-fit case is a challenge: this approach is distinct from the previous section in which a circular orbit was assumed.  For a circular orbit, we could use the ``flattening'' of the loop as a constraint on viewing inclination, under the assumption that the data capture nearly a full loop as representing half the period.  For elliptical orbits, there is a greater diversity of possibilities.  

First, we ignore the scaling factor $\tau_\ast/(1-e^2)^2$. Once a match to the data are obtained, $e$ will be constrained, and $\tau_\ast$ implied. Second, we now consider light curves in total polarization, which from eqs.~(\ref{eq:pol}),~(\ref{eq:dqu}),~and~(\ref{eq:dquu}) follows as $\Delta p = \sqrt{\Delta q^2 + \Delta u^2}$. In positive-definite polarization, we can match the characteristics of the data assuming the orbit was observed around periastron passage.  It is not the absolute values of polarization that matters most.  Instead, it is the fact that the observations display a double peak. The two peaks are mostly equal, and ratio of the peak to the trough between is about 1.5.  These characteristics help constrain the eccentricity of the orbit and the longitude of periastron passage.  The observed polarization is then scaled to match the pertinent orbital phases of the model.  Once achieved, the orbital period is implied.  

There is one additional parameter. The above deals strictly with the total polarization. Next is whether the model is consistent with the $\Delta q$ and $\Delta u$ measurements. In this regard, the best model is compared with observations in a $Q-U$ diagram. There remains one other free parameter which is orientation of the loop configuration on the sky. This relates to a rotation of the projected binary orbit on the sky in relation to how Stokes parameters are defined by the instrument.

\begin{figure*}
    \centering
    \includegraphics[trim={0.5em 0 12em 0},clip, width=4.5in]{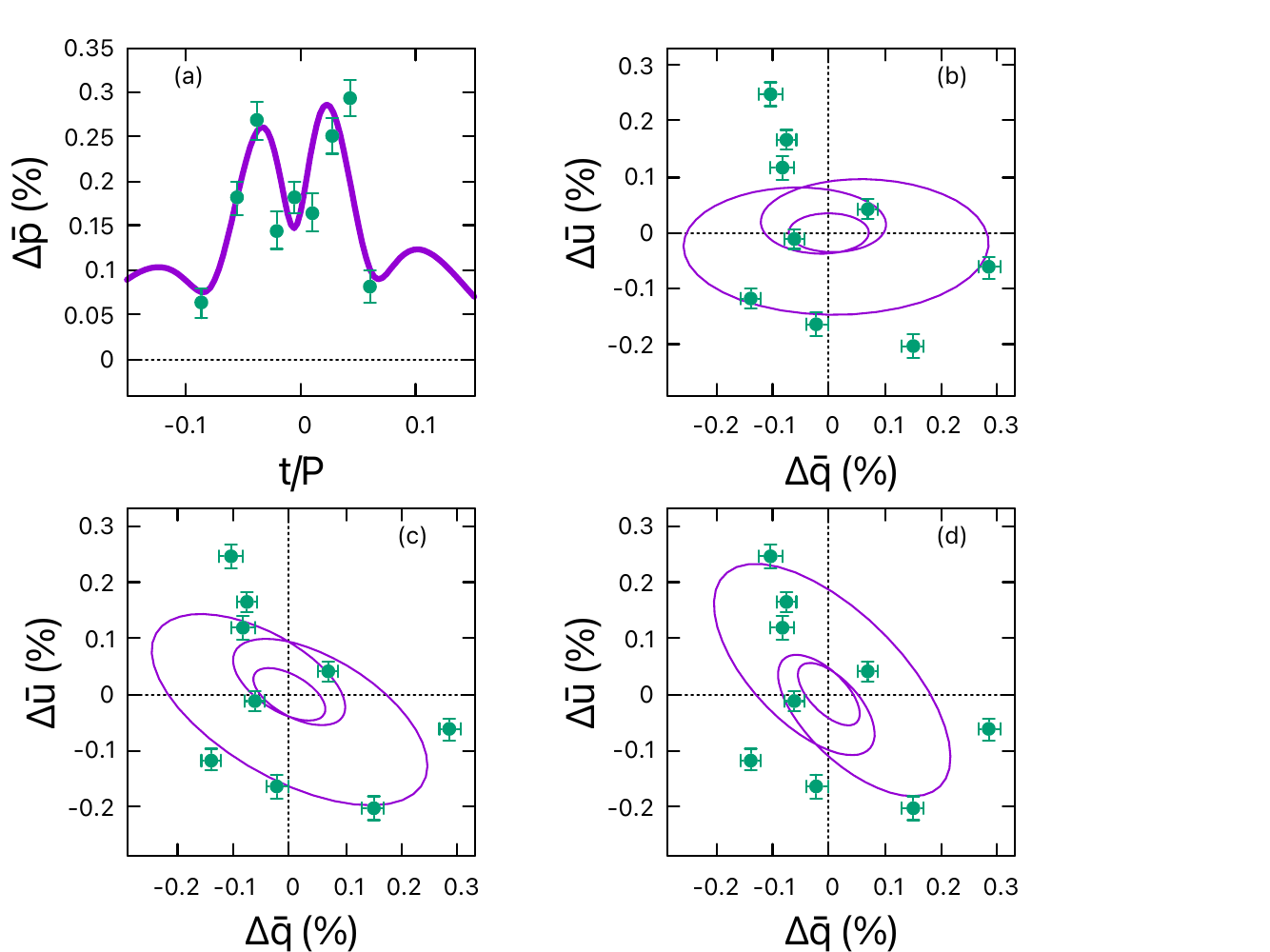}
    \caption{Comparison of the observations with an elliptical model that has $e=0.5$, $i=75^\circ$, and $\phi_{\rm p} = 140^\circ$.  (a) The bold purple curve is for $\Delta \bar{p}$.  Panels (b), (c), and (d) are results shown in $\Delta \bar{q}$ vs.\ $\Delta \bar{u}$ for $\psi=0^\circ$, $30^\circ$, and $225^\circ$, respectively.  The data points are weighted averages for all 4 passbands taken in a given night.}
    \label{fig:ellipfit}
\end{figure*}

\cite{1982MNRAS.198..787B} derived the expressions that we adopt to model the observations of \target; however, those authors did not provide many examples for model light curves or $Q-U$ diagrams.  In Figures \ref{fig:ecc0}-\ref{fig:ecc3}, we show a grid of models for different values of $e$ and $\phi_{\rm p}$, with inclination fixed at $i=75^\circ$.  A fairly high inclination is required to obtain the observed two peaks around periastron passage in the polarization light curve, and this inclination gives examples of eccentric orbits that can produce the peak-to-trough ratios of about 1.5.

Each these figures displays four panels, with orbital phase being time relative to the period, and eccentricities $e=0.0$ to 0.5 in steps of 0.1 for the fixed curves (see figure caption for colors).  The upper left panel is for $\Delta q$, the upper right panel for $\Delta u$, the lower right panel for $\Delta p = \sqrt{\Delta q^2 + \Delta u^2}$, and finally the lower left panel is the $\Delta q - \Delta u$ diagram.  For the latter, even though the two axes have the same ranges, the panel is not square, so the loops have a distorted appearance.  Figures \ref{fig:ecc0}-\ref{fig:ecc3} are for $\phi_{\rm p} =0^\circ, 130^\circ, 140^\circ$, and $150^\circ$.  Note that Figure~\ref{fig:ecc0} shows 3 peaks near the periastron passage.  The ratio between peaks is a function of $e$.  By contrast, the larger values of $\phi_{\rm p} = 130^\circ-150^\circ$ tend to show 2 peaks.  However, in Figure~\ref{fig:ecc2} the ratio of peaks is fairly equal with $e$.  In Figure~\ref{fig:ecc3}, the leading peak is the larger one, whereas for Figure~\ref{fig:ecc1}, the leading peak is the smaller one.  The observations of \target{} show fairly equal peaks.

The upper left panel of Figure~\ref{fig:ellipfit} shows a reasonable match between the observations for total polarization between the observations as points and the model light curve.  Note that for this figure, the four UBVR passband measurements taken on a given day have been averaged, reducing the 36 measurements to 9, as was done in the lower panel of Figure~\ref{fig:qufits}. The model is for $i=75^\circ$, $e=0.5$, and $\phi_{\rm p} = 140^\circ$. The nine days of observations span roughly an interval of 0.15 in orbital phase in the model, indicating an orbital period of about 70~d.

The other three panels are $Q-U$ diagrams where the model loop is shown at different orientations and compared to the data. The upper right panel (b) is with no rotation, so the loop is oriented in the idealized polarization frame defined by the star. Panel (d) at lower right is for a Mueller matrix rotation of $+225^\circ$; Panel (c) at lower left is for a rotation of $+30^\circ$. The polarization light curve is invariant to this rotation. By eye the best match appears to be in Panel~(c). Taking the orbital period as 70~d, the implied semi-major axis of the orbit is $5.8R_\ast$. With $e\sim 0.5$, the periastron and apastron distances are about $\sim 3R_\ast$ and $\sim 9R_\ast$, which respectively correspond to an angular separation of about 0.06~mas and 0.2~mas, using the stellar parameters from Table~\ref{tab:hen3-519_properties}. With recent advances in speckle imaging (for example, using the `Alopeke or Zorro cameras mounted on the Gemini-N/S telescopes, respectively) and long-baseline interferometry (for example, using the Center for High Angular Resolution Astronomy array), such sensitive measurements are now possible for the first time \citep{Shara2022,Richardson2024}. 

In summary, it is not surprising that a model for an elliptical orbit could be found to match the observations.  There are more free parameters as compared with the case of a circular orbit.  Given the limited sampling and uncertainties in the measurements, it is unreasonable to attempt a formal best fit to the data.  The match indicated in Figure~\ref{fig:ellipfit} is unlikely to be unique; however, the exercise is valuable in terms of identifying features that can help constrain a best-fit model.


\section{Discussion and Conclusions \label{sec:concl}}

Multi-color photopolarimetric observations were carried out for \target{}, producing nine measurements per (UBVR) filter over 10 nights. Each filter independently traces a loop in the $Q-U$ diagram and, after shifting the polarization through subtracting each filter's respective mean value, all four loops overlay quite well. This is exactly what is expected from polarization arising from electron scattering, which is a gray opacity. We interpret the loop as suggesting that \target{} has a binary companion. 

As a first test of the binary interpretation, we assume a circular orbit and apply the optically thin scattering result from \citet{1978A&A....68..415B}. The best fit provides an orbital period of 16.7~$\pm0.3$~d and an orbital radius of $3.2R_\ast$, assuming the companion star is the primary, and likely of spectral type O5-7, in accordance with other WN binary systems that have been previously characterized \citep{Niemela1995,Niemela1996,Muntean2009,Collado2013}. 
The uncertainties in the measurements do not allow a precise determination of the viewing inclination, except that the system is neither quite pole-on nor edge-on. The fit assuming a circular orbit suggests a fairly high inclination of $i \sim 80^{\circ}$. 

We also consider an elliptical orbit using the simplified model of \cite{1982MNRAS.198..787B}. The expectation is that the observations caught the companion around periastron passage, for which polarization achieves its maximum value. Unfortunately, the limited numbers of measurements combined with uncertainties in the values means that the elliptical case has simply too many free parameters to provide for a unique fit by the model. As in the case of a circular orbit, an edge-on viewing inclination is ruled out. A reasonable match to the data was obtained assuming a viewing inclination of $i\sim 75^\circ$ with a fairly eccentric orbit of $e\sim 0.5$.  This leads to an estimated orbital period of $\sim 70$~d and semi-major axis $a \sim 6R_\ast$, with periastron and apastron distances of about $3R_\ast$ and $9R_\ast$, corresponding to angular separations of 0.06 and 0.2~mas, respectively.

\target{} was included in the recent long-baseline interferometric survey of \citet{2024A&A...692A.109D}, which did not detect a companion; however, given the sensitivity of the survey, it is reasonable that a companion with an orbital period of less than 100~d may not have been within the detection limits of the program. The authors also point out that the binarity fraction for WNh stars is about 23\%, roughly half of what is predicted for cWR stars, and potentially relevant to this work, since \target{} displays Hydrogen in its spectrum \citep{Smith94}.
The spectroscopic multiplicity survey of late-type WN stars by \citet{2023A&A...674A..88D} did not include \target, but together with their earlier work reported in \citet{2022A&A...664A..93D} found that the orbital periods of Galactic WN binary systems are typically short, on the order of $<10$~d. A similar study of Galactic WC stars found those systems are expected to have much longer orbital periods of 100~d or more \citep{2020A&A...641A..26D}. While new observations of \target{} will be necessary to confirm whether a binary companion is indeed present (and if so, what the nature of the companion is, and whether the WN star is the primary or secondary of the system), the results of our models are generally consistent with these expectations, and can provide guidance for future observational planning. 

An alternative to the binary interpretation could be a co-rotating interaction region (CIR). These are asymmetric flow structures in the stellar wind that take on spiral morphologies and co-rotate with the star \citep{1984ApJ...283..303M}. CIRs have been associated with the ubiquitous discrete absorption components, or DACs, seen in the UV lines of many OB stars \citep{1986ApJS...61..357P, 1996A&AS..116..257K}. The physical cause for producing these wind structures is typically associated with stellar bright spots \citep{1996ApJ...462..469C, 2008ApJ...678..408L}, which has been confirmed in the heavily studied case of the O4I star, $\zeta$ Pup \citep{2018MNRAS.473.5532R}. Among the WR stars, there have been reports for CIRs in several WN-type stars \citep{2009ApJ...698.1951S, 2011ApJ...736..140C}. A handful of studies have explored the variable polarization from CIRs \citep{2015A&A...575A.129I, 2019MNRAS.489.2873C}, providing useful comparisons with the binary hypothesis.  

An equatorial CIR and a circular binary orbit produce similar morphologies in $Q-U$ diagrams, specifically two loops per rotation or per orbit, respectively. And CIRs can produce a polarization amplitude comparable to that of a binary scenario. For example, \cite{2025RNAAS...9...36S} has reported on variable polarization in WR~1 (HD~4004) over about 5 weeks of observations. This star had previously been reported as a candidate for a CIR \citep{2011ApJ...736..140C}.  \cite{2025RNAAS...9...36S} find two consecutive loops in the $Q-U$ diagram that match the previously reported 17~d period. The scale of variable polarization is similar to that reported here for \target.

However, a major difference is that a binary companion is spatially removed from the primary star, whereas the CIR structure extends down to the starspot. While the current models for variable polarization allow for multiple scattering appropriate to WR wind densities, the models lack absorptive opacities (e.g., free-free and bound-free). The high mass-loss rate and slow wind speed for \target{} suggests high density and absorptive optical depths that should alter the loop morphologies of CIRs in $Q-U$ diagrams relative to existing predictions.  

Comparison between variable polarization for a CIR and an elliptical orbit is quite distinct. Depending on the latitude of the CIR's origin combined with viewing inclination, it is possible to produce either one or two loops per rotation period \citep[e.g.,][]{2015A&A...575A.129I}; however, it corotates. By contrast, for a binary elliptical orbit with an eccentricity of a couple tenths or more, variable polarization is maximized near periastron passage and minimized around apastron. So while there can be two or more loops based on \cite{1982MNRAS.198..787B}, considerably less time is spent between the loops. For the case of \target, our application of an elliptical orbit would suggest that the high polarization state takes about one-seventh of the 70~d orbit, but that the lower polarization state would take the other two thirds. Such a duty cycle is not something that would occur for a CIR. A new time-resolved polarimetric study of \target{} over the proposed orbital period may be able to distinguish between a CIR or binary companion scenario.

Another distinction between binaries and CIRs is that binaries have a well-defined ephemeris. By contrast, CIRs maintain periods because of rotation but show phase drifts. The coherence time of a CIR structure is weeks or months \citep[e.g., ][]{2011ApJ...736..140C, 2016MNRAS.460.3407A}. A weak argument against a CIR in \target{} is that CIRs have so far only been claimed in H-deficient broad-lined WN stars, specifically WR~1, WR~6, WR~110, and WR~134 \citep{2011ApJ...735...34C,2016MNRAS.460.3407A,2018MNRAS.474.1886S,2023MNRAS.526.1298I,2025RNAAS...9...36S} 
By contrast \target{} is late WN star, has Hydrogen in its spectrum, and its lines are not broad \citep{Smith94}.  While there are not many CIR candidates among the WR stars, if confirmed in \target, it would be a deviation from the trend so far.  

It would also be a deviation from the other examples in terms of rotation speed. The four WR stars with CIRs have rotation periods on the order of days, but their hydrostatic radii are much more compact at about $1-2R_\odot$ than \target{} (c.f., Tab.~\ref{tab:hen3-519_properties}).  Assuming an Eddington parameter $\Gamma=0$, those stars would be rotating at a few percent of critical break-up speed. The Q-U loops for CIRs based on thin scattering models can have one or two loops per rotation. If \target{} has a CIR, the rotation period could be 8.5 or 17 days. The rotation speed at the equator would be 190 or 95~\kms, respectively. Assuming again that $\Gamma=0$, the critical rotation for \target{} is about 320~\kms, so a CIR would suggest a rotation of 30-60\% of critical.  This is quite fast but would be consistent with suggestions by \cite{2011A&A...536L..10V} that post-LBVs can retain significant spin to act as progenitors of gamma-ray bursts (see also \citealt{2013A&A...558A.131G} for a similar discussion regarding WO stars).

While the data are suggestive and the orbital solutions tantalizing, new observations are clearly needed. Foremost is the need to determine whether a loop persists, as indicated in the data. For the binary interpretation to remain plausible, the loop must recur cyclically. It is important to confirm the loop persists and to obtain two complete cycles of observations with detailed phase sampling in order to distinguish between the circular and elliptical orbit scenarios. Such a study would enable improved constraints on the orbital inclination (assuming the binary hypothesis is confirmed) and inform future radial velocity searches from which the stellar masses could be refined.

It should be noted that massive stars have structured winds. The WR stars can have quite clumpy wind flows \citep[e.g.,][]{2023MNRAS.519.3271I}. These clumps produce stochastically variable polarization that could lead to loops that are not entirely identical from cycle to cycle. In fact, such an effect is seen in WR~1:  the two consecutive loops complete in the same time of 17~d but are not consistent with having the same shape, likely a consequence of clumping effects \citep{2025RNAAS...9...36S}. 

Studies of individual LBV or post-LBV objects like \target{} contribute to an evolving picture of massive stellar evolution, revealing the effect of strong winds or of a binary companion in these influential post-Main Sequence stages. Massive stars are generally understood to be born in multiple systems \citep{Sana2012,Sana2014,Moe2017}. If \target{} is a post-LBV object, hosting a binary companion in a fairly modest orbit could suggest orbital degradation from an earlier evolutionary phase when the primary was larger. After all, at much larger sizes of the primary, the orbit from our analysis would place the secondary inside the stellar atmosphere of the primary. Related, if \target{} is evolving to become a ``classical'' Wolf-Rayet star \citep{1996MNRAS.281..163S}, then the binary orbit will become relatively wider (in the sense of stellar radii) as \target{} becomes smaller in size.  We note that there are several WR binaries with close companions in short periods, including CQ~Cep (1.6~d), CX~Cep (2.1~d), and V444~Cyg (4.2~d) \citep{2023MNRAS.523.1524S}.
Verification of binarity or CIR activity in \target{} helps address the pressing current questions about the evolutionary diversity among massive star multiples and/or their rotation periods \citep{2024ARA&A..62...21M}.  


\begin{acknowledgements}
The authors wish to thankfully acknowledge thoughtful contributions from Jorick Vink and V{\'e}ronique Petit that provided guidance for this work. We also thank the referee for their careful reading and helpful suggestions.
CE gratefully acknowledges financial support for travel related to this work provided by the STScI Director's Research Funds.
FS gratefully acknowledges support from the East Tennessee State University Office of Undergraduate Research and Creative Activities and the Ronald E. McNair Postbaccalaureate Achievement Program, a TRIO program that sponsored participation in the Advanced Research Internship which contributed to this work.
\end{acknowledgements}

\facilities{
Based on observations collected at the European Southern Observatory under ESO program 078.D-0790(A) (PI: Davies). This research also referenced the SVO Filter Profile Service ``Carlos Rodrigo'' \citep{SVO-1,SVO-2,SVO-3}, funded by MCIN/AEI/10.13039/501100011033/ through grant PID2023-146210NB-I00.
}

\software{
\texttt{Astropy} \citep{astropy, astropy2, astropy3}, 
\texttt{Matplotlib} \citep{matplotlib}, 
\texttt{NumPy} \citep{numpy, numpy2},
\texttt{emcee} \citep{Foreman-Mackey_2013},
\texttt{Gnuplot} \citep{gnuplot}
}

\bibliographystyle{aasjournal}
\bibliography{lbvpol}

\begin{thebibliography}{}
\expandafter\ifx\csname natexlab\endcsname\relax\def\natexlab#1{#1}\fi
\providecommand{\url}[1]{\href{#1}{#1}}
\providecommand{\dodoi}[1]{doi:~\href{http://doi.org/#1}{\nolinkurl{#1}}}
\providecommand{\doeprint}[1]{\href{http://ascl.net/#1}{\nolinkurl{http://ascl.net/#1}}}
\providecommand{\doarXiv}[1]{\href{https://arxiv.org/abs/#1}{\nolinkurl{https://arxiv.org/abs/#1}}}

\bibitem[{M. {Aghakhanloo} {et~al.}(2017){Aghakhanloo}, {Murphy}, {Smith}, \& {Hlo{\v{z}}ek}}]{2017MNRAS.472..591A}
{Aghakhanloo}, M., {Murphy}, J.~W., {Smith}, N., \& {Hlo{\v{z}}ek}, R. 2017, \bibinfo{title}{{Modelling luminous-blue-variable isolation},} \mnras, 472, 591, \dodoi{10.1093/mnras/stx2050}

\bibitem[{M. {Aghakhanloo} {et~al.}(2022){Aghakhanloo}, {Smith}, {Andrews}, {Olsen}, {Besla}, \& {Choi}}]{2022MNRAS.516.2142A}
{Aghakhanloo}, M., {Smith}, N., {Andrews}, J., {et~al.} 2022, \bibinfo{title}{{Kinematics of luminous blue variables in the Large Magellanic Cloud},} \mnras, 516, 2142, \dodoi{10.1093/mnras/stac2265}

\bibitem[{C. {Agliozzo} {et~al.}(2021){Agliozzo}, {Phillips}, {Mehner}, {Baade}, {Scicluna}, {Kemper}, {Asmus}, {de Wit}, \& {Pignata}}]{Agliozzo2021}
{Agliozzo}, C., {Phillips}, N., {Mehner}, A., {et~al.} 2021, \bibinfo{title}{{The contribution by luminous blue variable stars to the dust content of the Magellanic Clouds},} \aap, 655, A98, \dodoi{10.1051/0004-6361/202141279}

\bibitem[{E.~J. {Aldoretta} {et~al.}(2016){Aldoretta}, {St-Louis}, {Richardson}, {Moffat}, {Eversberg}, {Hill}, {Shenar}, {Artigau}, {Gauza}, {Knapen}, {Kub{\'a}t}, {Kub{\'a}tov{\'a}}, {Maltais-Tariant}, {Mu{\~n}oz}, {Pablo}, {Ramiaramanantsoa}, {Richard-Laferri{\`e}re}, {Sablowski}, {Sim{\'o}n-D{\'\i}az}, {St-Jean}, {Bolduan}, {Dias}, {Dubreuil}, {Fuchs}, {Garrel}, {Grutzeck}, {Hunger}, {K{\"u}sters}, {Langenbrink}, {Leadbeater}, {Li}, {Lopez}, {Mauclaire}, {Moldenhawer}, {Potter}, {dos Santos}, {Schanne}, {Schmidt}, {Sieske}, {Strachan}, {Stinner}, {Stinner}, {Stober}, {Strandbaek}, {Syder}, {Verilhac}, {Waldschl{\"a}ger}, {Weiss}, \& {Wendt}}]{2016MNRAS.460.3407A}
{Aldoretta}, E.~J., {St-Louis}, N., {Richardson}, N.~D., {et~al.} 2016, \bibinfo{title}{{An extensive spectroscopic time series of three Wolf-Rayet stars - I. The lifetime of large-scale structures in the wind of WR 134},} \mnras, 460, 3407, \dodoi{10.1093/mnras/stw1188}

\bibitem[{ {Astropy Collaboration} {et~al.}(2013){Astropy Collaboration}, {Robitaille}, {Tollerud}, {Greenfield}, {Droettboom}, {Bray}, {Aldcroft}, {Davis}, {Ginsburg}, {Price-Whelan}, {Kerzendorf}, {Conley}, {Crighton}, {Barbary}, {Muna}, {Ferguson}, {Grollier}, {Parikh}, {Nair}, {Unther}, {Deil}, {Woillez}, {Conseil}, {Kramer}, {Turner}, {Singer}, {Fox}, {Weaver}, {Zabalza}, {Edwards}, {Azalee Bostroem}, {Burke}, {Casey}, {Crawford}, {Dencheva}, {Ely}, {Jenness}, {Labrie}, {Lim}, {Pierfederici}, {Pontzen}, {Ptak}, {Refsdal}, {Servillat}, \& {Streicher}}]{astropy}
{Astropy Collaboration}, {Robitaille}, T.~P., {Tollerud}, E.~J., {et~al.} 2013, \bibinfo{title}{{Astropy: A community Python package for astronomy},} \aap, 558, A33, \dodoi{10.1051/0004-6361/201322068}

\bibitem[{ {Astropy Collaboration} {et~al.}(2018){Astropy Collaboration}, {Price-Whelan}, {Sip{\H{o}}cz}, {G{\"u}nther}, {Lim}, {Crawford}, {Conseil}, {Shupe}, {Craig}, {Dencheva}, {Ginsburg}, {VanderPlas}, {Bradley}, {P{\'e}rez-Su{\'a}rez}, {de Val-Borro}, {Aldcroft}, {Cruz}, {Robitaille}, {Tollerud}, {Ardelean}, {Babej}, {Bach}, {Bachetti}, {Bakanov}, {Bamford}, {Barentsen}, {Barmby}, {Baumbach}, {Berry}, {Biscani}, {Boquien}, {Bostroem}, {Bouma}, {Brammer}, {Bray}, {Breytenbach}, {Buddelmeijer}, {Burke}, {Calderone}, {Cano Rodr{\'\i}guez}, {Cara}, {Cardoso}, {Cheedella}, {Copin}, {Corrales}, {Crichton}, {D'Avella}, {Deil}, {Depagne}, {Dietrich}, {Donath}, {Droettboom}, {Earl}, {Erben}, {Fabbro}, {Ferreira}, {Finethy}, {Fox}, {Garrison}, {Gibbons}, {Goldstein}, {Gommers}, {Greco}, {Greenfield}, {Groener}, {Grollier}, {Hagen}, {Hirst}, {Homeier}, {Horton}, {Hosseinzadeh}, {Hu}, {Hunkeler}, {Ivezi{\'c}}, {Jain}, {Jenness}, {Kanarek}, {Kendrew}, {Kern}, {Kerzendorf}, {Khvalko}, {King}, {Kirkby}, {Kulkarni},
  {Kumar}, {Lee}, {Lenz}, {Littlefair}, {Ma}, {Macleod}, {Mastropietro}, {McCully}, {Montagnac}, {Morris}, {Mueller}, {Mumford}, {Muna}, {Murphy}, {Nelson}, {Nguyen}, {Ninan}, {N{\"o}the}, {Ogaz}, {Oh}, {Parejko}, {Parley}, {Pascual}, {Patil}, {Patil}, {Plunkett}, {Prochaska}, {Rastogi}, {Reddy Janga}, {Sabater}, {Sakurikar}, {Seifert}, {Sherbert}, {Sherwood-Taylor}, {Shih}, {Sick}, {Silbiger}, {Singanamalla}, {Singer}, {Sladen}, {Sooley}, {Sornarajah}, {Streicher}, {Teuben}, {Thomas}, {Tremblay}, {Turner}, {Terr{\'o}n}, {van Kerkwijk}, {de la Vega}, {Watkins}, {Weaver}, {Whitmore}, {Woillez}, {Zabalza}, \& {Astropy Contributors}}]{astropy2}
{Astropy Collaboration}, {Price-Whelan}, A.~M., {Sip{\H{o}}cz}, B.~M., {et~al.} 2018, \bibinfo{title}{{The Astropy Project: Building an Open-science Project and Status of the v2.0 Core Package},} \aj, 156, 123, \dodoi{10.3847/1538-3881/aabc4f}

\bibitem[{ {Astropy Collaboration} {et~al.}(2022){Astropy Collaboration}, {Price-Whelan}, {Lim}, {Earl}, {Starkman}, {Bradley}, {Shupe}, {Patil}, {Corrales}, {Brasseur}, {N{\"o}the}, {Donath}, {Tollerud}, {Morris}, {Ginsburg}, {Vaher}, {Weaver}, {Tocknell}, {Jamieson}, {van Kerkwijk}, {Robitaille}, {Merry}, {Bachetti}, {G{\"u}nther}, {Aldcroft}, {Alvarado-Montes}, {Archibald}, {B{\'o}di}, {Bapat}, {Barentsen}, {Baz{\'a}n}, {Biswas}, {Boquien}, {Burke}, {Cara}, {Cara}, {Conroy}, {Conseil}, {Craig}, {Cross}, {Cruz}, {D'Eugenio}, {Dencheva}, {Devillepoix}, {Dietrich}, {Eigenbrot}, {Erben}, {Ferreira}, {Foreman-Mackey}, {Fox}, {Freij}, {Garg}, {Geda}, {Glattly}, {Gondhalekar}, {Gordon}, {Grant}, {Greenfield}, {Groener}, {Guest}, {Gurovich}, {Handberg}, {Hart}, {Hatfield-Dodds}, {Homeier}, {Hosseinzadeh}, {Jenness}, {Jones}, {Joseph}, {Kalmbach}, {Karamehmetoglu}, {Ka{\l}uszy{\'n}ski}, {Kelley}, {Kern}, {Kerzendorf}, {Koch}, {Kulumani}, {Lee}, {Ly}, {Ma}, {MacBride}, {Maljaars}, {Muna}, {Murphy}, {Norman},
  {O'Steen}, {Oman}, {Pacifici}, {Pascual}, {Pascual-Granado}, {Patil}, {Perren}, {Pickering}, {Rastogi}, {Roulston}, {Ryan}, {Rykoff}, {Sabater}, {Sakurikar}, {Salgado}, {Sanghi}, {Saunders}, {Savchenko}, {Schwardt}, {Seifert-Eckert}, {Shih}, {Jain}, {Shukla}, {Sick}, {Simpson}, {Singanamalla}, {Singer}, {Singhal}, {Sinha}, {Sip{\H{o}}cz}, {Spitler}, {Stansby}, {Streicher}, {{\v{S}}umak}, {Swinbank}, {Taranu}, {Tewary}, {Tremblay}, {Val-Borro}, {Van Kooten}, {Vasovi{\'c}}, {Verma}, {de Miranda Cardoso}, {Williams}, {Wilson}, {Winkel}, {Wood-Vasey}, {Xue}, {Yoachim}, {Zhang}, {Zonca}, \& {Astropy Project Contributors}}]{astropy3}
{Astropy Collaboration}, {Price-Whelan}, A.~M., {Lim}, P.~L., {et~al.} 2022, \bibinfo{title}{{The Astropy Project: Sustaining and Growing a Community-oriented Open-source Project and the Latest Major Release (v5.0) of the Core Package},} \apj, 935, 167, \dodoi{10.3847/1538-4357/ac7c74}

\bibitem[{J.~C. {Brown} {et~al.}(1982){Brown}, {Aspin}, {Simmons}, \& {McLean}}]{1982MNRAS.198..787B}
{Brown}, J.~C., {Aspin}, C., {Simmons}, J.~F.~L., \& {McLean}, I.~S. 1982, \bibinfo{title}{{The effect of orbital eccentricity on polarimetric binary diagnostics.},} \mnras, 198, 787, \dodoi{10.1093/mnras/198.3.787}

\bibitem[{J.~C. {Brown} {et~al.}(1978){Brown}, {McLean}, \& {Emslie}}]{1978A&A....68..415B}
{Brown}, J.~C., {McLean}, I.~S., \& {Emslie}, A.~G. 1978, \bibinfo{title}{{Polarisation by Thomson scattering in optically thin stellar envelopes. II. Binary and multiple star envelopes and the determination of binary inclinations.},} \aap, 68, 415

\bibitem[{D. {Carlos-Leblanc} {et~al.}(2019){Carlos-Leblanc}, {St-Louis}, {Bjorkman}, \& {Ignace}}]{2019MNRAS.489.2873C}
{Carlos-Leblanc}, D., {St-Louis}, N., {Bjorkman}, J.~E., \& {Ignace}, R. 2019, \bibinfo{title}{{Monte Carlo simulations of polarimetric and light variability from corotating interaction regions in hot stellar winds},} \mnras, 489, 2873, \dodoi{10.1093/mnras/stz2273}

\bibitem[{A.~N. {Chen{\'e}} \& N. {St-Louis}(2011){Chen{\'e}} \& {St-Louis}}]{2011ApJ...736..140C}
{Chen{\'e}}, A.~N., \& {St-Louis}, N. 2011, \bibinfo{title}{{A Systematic Search for Corotating Interaction Regions in Apparently Single Galactic Wolf-Rayet Stars. II. A Global View of the Wind Variability},} \apj, 736, 140, \dodoi{10.1088/0004-637X/736/2/140}

\bibitem[{A.~N. {Chen{\'e}} {et~al.}(2011){Chen{\'e}}, {Moffat}, {Cameron}, {Fahed}, {Gamen}, {Lef{\`e}vre}, {Rowe}, {St-louis}, {Muntean}, {De La Chevroti{\`e}re}, {Guenther}, {Kuschnig}, {Matthews}, {Rucinski}, {Sasselov}, \& {Weiss}}]{2011ApJ...735...34C}
{Chen{\'e}}, A.~N., {Moffat}, A.~F.~J., {Cameron}, C., {et~al.} 2011, \bibinfo{title}{{WR 110: A Single Wolf-Rayet Star with Corotating Interaction Regions in its Wind?},} \apj, 735, 34, \dodoi{10.1088/0004-637X/735/1/34}

\bibitem[{M. {Clampin} {et~al.}(1995){Clampin}, {Schulte-Ladbeck}, {Nota}, {Robberto}, {Paresce}, \& {Clayton}}]{1995AJ....110..251C}
{Clampin}, M., {Schulte-Ladbeck}, R.~E., {Nota}, A., {et~al.} 1995, \bibinfo{title}{{High Resolution Coronographic Imaging and Spectropolarimetry of the HR Carinae Nebula},} \aj, 110, 251, \dodoi{10.1086/117514}

\bibitem[{D. {Clarke}(2010){Clarke}}]{2010stpo.book.....C}
{Clarke}, D. 2010, {Stellar Polarimetry} (Wiley-VCH Verlag GmbH \& Co. KGaA)

\bibitem[{ {Collado, A.} {et~al.}(2013){Collado, A.}, {Gamen, R.}, \& {Barbá, R. H.}}]{Collado2013}
{Collado, A.}, {Gamen, R.}, \& {Barbá, R. H.} 2013, \bibinfo{title}{The new Wolf-Rayet binary system WR62a,} A\&A, 552, A22, \dodoi{10.1051/0004-6361/201118460}

\bibitem[{S.~R. {Cranmer} \& S.~P. {Owocki}(1996){Cranmer} \& {Owocki}}]{1996ApJ...462..469C}
{Cranmer}, S.~R., \& {Owocki}, S.~P. 1996, \bibinfo{title}{{Hydrodynamical Simulations of Corotating Interaction Regions and Discrete Absorption Components in Rotating O-Star Winds},} \apj, 462, 469, \dodoi{10.1086/177166}

\bibitem[{K. {Davidson} {et~al.}(1993){Davidson}, {Humphreys}, {Hajian}, \& {Terzian}}]{Davidson93}
{Davidson}, K., {Humphreys}, R.~M., {Hajian}, A., \& {Terzian}, Y. 1993, \bibinfo{title}{{He 3-519: A Peculiar Post-LBV, Pre-WN Star?},} \apj, 411, 336, \dodoi{10.1086/172833}

\bibitem[{B. {Davies} {et~al.}(2005){Davies}, {Oudmaijer}, \& {Vink}}]{2005A&A...439.1107D}
{Davies}, B., {Oudmaijer}, R.~D., \& {Vink}, J.~S. 2005, \bibinfo{title}{{Asphericity and clumpiness in the winds of Luminous Blue Variables},} \aap, 439, 1107, \dodoi{10.1051/0004-6361:20052781}

\bibitem[{K. {Deshmukh} {et~al.}(2024){Deshmukh}, {Sana}, {M{\'e}rand}, {Bordier}, {Langer}, {Bodensteiner}, {Dsilva}, {Frost}, {Gosset}, {Le Bouquin}, {Lefever}, {Mahy}, {Patrick}, {Reggiani}, {Sander}, {Shenar}, {Tramper}, {Villase{\~n}or}, \& {Waisberg}}]{2024A&A...692A.109D}
{Deshmukh}, K., {Sana}, H., {M{\'e}rand}, A., {et~al.} 2024, \bibinfo{title}{{Investigating 39 Galactic Wolf-Rayet stars with VLTI/GRAVITY: Uncovering a long-period binary desert},} \aap, 692, A109, \dodoi{10.1051/0004-6361/202452352}

\bibitem[{K. {Dsilva} {et~al.}(2020){Dsilva}, {Shenar}, {Sana}, \& {Marchant}}]{2020A&A...641A..26D}
{Dsilva}, K., {Shenar}, T., {Sana}, H., \& {Marchant}, P. 2020, \bibinfo{title}{{A spectroscopic multiplicity survey of Galactic Wolf-Rayet stars. I. The northern WC sequence},} \aap, 641, A26, \dodoi{10.1051/0004-6361/202038446}

\bibitem[{K. {Dsilva} {et~al.}(2022){Dsilva}, {Shenar}, {Sana}, \& {Marchant}}]{2022A&A...664A..93D}
{Dsilva}, K., {Shenar}, T., {Sana}, H., \& {Marchant}, P. 2022, \bibinfo{title}{{A spectroscopic multiplicity survey of Galactic Wolf-Rayet stars. II. The northern WNE sequence},} \aap, 664, A93, \dodoi{10.1051/0004-6361/202142729}

\bibitem[{K. {Dsilva} {et~al.}(2023){Dsilva}, {Shenar}, {Sana}, \& {Marchant}}]{2023A&A...674A..88D}
{Dsilva}, K., {Shenar}, T., {Sana}, H., \& {Marchant}, P. 2023, \bibinfo{title}{{A spectroscopic multiplicity survey of Galactic Wolf-Rayet stars . III. The northern late-type nitrogen-rich sample},} \aap, 674, A88, \dodoi{10.1051/0004-6361/202244308}

\bibitem[{T. {Eversberg} {et~al.}(1999){Eversberg}, {Moffat}, \& {Marchenko}}]{1999PASP..111..861E}
{Eversberg}, T., {Moffat}, A.~F.~J., \& {Marchenko}, S.~V. 1999, \bibinfo{title}{{Spectropolarimetry of the WR+O Binary gamma\^2 Velorum during Periastron Passage},} \pasp, 111, 861, \dodoi{10.1086/316390}

\bibitem[{D. Foreman-Mackey {et~al.}(2013)Foreman-Mackey, Hogg, Lang, \& Goodman}]{Foreman-Mackey_2013}
Foreman-Mackey, D., Hogg, D.~W., Lang, D., \& Goodman, J. 2013, \bibinfo{title}{emcee: The MCMC Hammer,} Publications of the Astronomical Society of the Pacific, 125, 306, \dodoi{10.1086/670067}

\bibitem[{A.~G. {Fullard} {et~al.}(2022){Fullard}, {O'Brien}, {Kerzendorf}, {Shrestha}, {Hoffman}, {Ignace}, \& {van der Smagt}}]{2022ApJ...930...89F}
{Fullard}, A.~G., {O'Brien}, J.~T., {Kerzendorf}, W.~E., {et~al.} 2022, \bibinfo{title}{{New Mass Estimates for Massive Binary Systems: A Probabilistic Approach Using Polarimetric Radiative Transfer},} \apj, 930, 89, \dodoi{10.3847/1538-4357/ac589e}

\bibitem[{K. {Gootkin} {et~al.}(2020){Gootkin}, {Dorn-Wallenstein}, {Lomax}, {Eadie}, {Levesque}, {Babler}, {Hoffman}, {Meade}, {Nordsieck}, \& {Wisniewski}}]{2020ApJ...900..162G}
{Gootkin}, K., {Dorn-Wallenstein}, T., {Lomax}, J.~R., {et~al.} 2020, \bibinfo{title}{{13 yr of P Cygni Spectropolarimetry: Investigating Mass Loss through H{\ensuremath{\alpha}}, Periodicity, and Ellipticity},} \apj, 900, 162, \dodoi{10.3847/1538-4357/abad32}

\bibitem[{J.~H. {Groh} {et~al.}(2013){Groh}, {Meynet}, {Georgy}, \& {Ekstr{\"o}m}}]{2013A&A...558A.131G}
{Groh}, J.~H., {Meynet}, G., {Georgy}, C., \& {Ekstr{\"o}m}, S. 2013, \bibinfo{title}{{Fundamental properties of core-collapse supernova and GRB progenitors: predicting the look of massive stars before death},} \aap, 558, A131, \dodoi{10.1051/0004-6361/201321906}

\bibitem[{W.~R. {Hamann} \& G. {Graefener}(2007){Hamann} \& {Graefener}}]{Hamann2007}
{Hamann}, W.~R., \& {Graefener}, G. 2007, in Astronomical Society of the Pacific Conference Series, Vol. 367, Massive Stars in Interactive Binaries, ed. N.~{St. -Louis} \& A.~F.~J. {Moffat}, 141

\bibitem[{P. {Harmanec}(2002){Harmanec}}]{2002AN....323...87H}
{Harmanec}, P. 2002, \bibinfo{title}{{The ever challenging emission-line binary beta Lyrae},} Astronomische Nachrichten, 323, 87, \dodoi{10.1002/1521-3994(200207)323:2<87::AID-ASNA87>3.0.CO;2-P}

\bibitem[{C.~R. {Harris} {et~al.}(2020){Harris}, {Millman}, {van der Walt}, {Gommers}, {Virtanen}, {Cournapeau}, {Wieser}, {Taylor}, {Berg}, {Smith}, {Kern}, {Picus}, {Hoyer}, {van Kerkwijk}, {Brett}, {Haldane}, {del R{\'\i}o}, {Wiebe}, {Peterson}, {G{\'e}rard-Marchant}, {Sheppard}, {Reddy}, {Weckesser}, {Abbasi}, {Gohlke}, \& {Oliphant}}]{numpy2}
{Harris}, C.~R., {Millman}, K.~J., {van der Walt}, S.~J., {et~al.} 2020, \bibinfo{title}{{Array programming with NumPy},} \nat, 585, 357, \dodoi{10.1038/s41586-020-2649-2}

\bibitem[{E.~R. {Higgins} {et~al.}(2021){Higgins}, {Sander}, {Vink}, \& {Hirschi}}]{2021MNRAS.505.4874H}
{Higgins}, E.~R., {Sander}, A.~A.~C., {Vink}, J.~S., \& {Hirschi}, R. 2021, \bibinfo{title}{{Evolution of Wolf-Rayet stars as black hole progenitors},} \mnras, 505, 4874, \dodoi{10.1093/mnras/stab1548}

\bibitem[{P.~F. {Hopkins} {et~al.}(2012){Hopkins}, {Quataert}, \& {Murray}}]{2012MNRAS.421.3522H}
{Hopkins}, P.~F., {Quataert}, E., \& {Murray}, N. 2012, \bibinfo{title}{{Stellar feedback in galaxies and the origin of galaxy-scale winds},} \mnras, 421, 3522, \dodoi{10.1111/j.1365-2966.2012.20593.x}

\bibitem[{R.~M. {Humphreys} \& K. {Davidson}(1994){Humphreys} \& {Davidson}}]{1994PASP..106.1025H}
{Humphreys}, R.~M., \& {Davidson}, K. 1994, \bibinfo{title}{{The Luminous Blue Variables: Astrophysical Geysers},} \pasp, 106, 1025, \dodoi{10.1086/133478}

\bibitem[{J.~D. Hunter(2007)Hunter}]{matplotlib}
Hunter, J.~D. 2007, \bibinfo{title}{Matplotlib: A 2D graphics environment,} Computing In Science \& Engineering, 9, 90

\bibitem[{R. {Ignace} {et~al.}(2023{\natexlab{a}}){Ignace}, {Bjorkman}, {Chen{\'e}}, {Erba}, {Fabiani}, {Moffat}, {Sincennes}, \& {St-Louis}}]{2023MNRAS.526.1298I}
{Ignace}, R., {Bjorkman}, J.~E., {Chen{\'e}}, A.~N., {et~al.} 2023{\natexlab{a}}, \bibinfo{title}{{Modelling variable linear polarization produced by Co-rotating Interaction Regions (CIRs) across optical recombination lines of Wolf-Rayet stars},} \mnras, 526, 1298, \dodoi{10.1093/mnras/stad2878}

\bibitem[{R. {Ignace} {et~al.}(2022){Ignace}, {Fullard}, {Shrestha}, {Naz{\'e}}, {Gayley}, {Hoffman}, {Lomax}, \& {St-Louis}}]{2022ApJ...933....5I}
{Ignace}, R., {Fullard}, A., {Shrestha}, M., {et~al.} 2022, \bibinfo{title}{{Modeling the Optical to Ultraviolet Polarimetric Variability from Thomson Scattering in Colliding-wind Binaries},} \apj, 933, 5, \dodoi{10.3847/1538-4357/ac6fce}

\bibitem[{R. {Ignace} {et~al.}(2025){Ignace}, {Gayley}, {Casini}, {Scowen}, {Erba}, \& {Drake}}]{2025arXiv250402659I}
{Ignace}, R., {Gayley}, K., {Casini}, R., {et~al.} 2025, \bibinfo{title}{{Spectropolarimetry for Discerning Geometry and Structure in Circumstellar Media of Hot Massive Stars},} arXiv e-prints, arXiv:2504.02659, \dodoi{10.48550/arXiv.2504.02659}

\bibitem[{R. {Ignace} {et~al.}(2023{\natexlab{b}}){Ignace}, {Moffat}, {Robert}, \& {Drissen}}]{2023MNRAS.519.3271I}
{Ignace}, R., {Moffat}, A.~F.~J., {Robert}, C., \& {Drissen}, L. 2023{\natexlab{b}}, \bibinfo{title}{{Constraints on clumps in the representative wind of the WN8 Wolf-Rayet star HD 96548 = WR 40 with simultaneous broad-band light and linear-polarization variability},} \mnras, 519, 3271, \dodoi{10.1093/mnras/stac3772}

\bibitem[{R. {Ignace} {et~al.}(2015){Ignace}, {St-Louis}, \& {Proulx-Giraldeau}}]{2015A&A...575A.129I}
{Ignace}, R., {St-Louis}, N., \& {Proulx-Giraldeau}, F. 2015, \bibinfo{title}{{Polarimetric modeling of corotating interaction regions threading massive-star winds},} \aap, 575, A129, \dodoi{10.1051/0004-6361/201424806}

\bibitem[{L. {Kaper} {et~al.}(1996){Kaper}, {Henrichs}, {Nichols}, {Snoek}, {Volten}, \& {Zwarthoed}}]{1996A&AS..116..257K}
{Kaper}, L., {Henrichs}, H.~F., {Nichols}, J.~S., {et~al.} 1996, \bibinfo{title}{{Long- and short-term variability in O-star winds. I. Time series of UV spectra for 10 bright O stars.},} \aaps, 116, 257

\bibitem[{R. {Klement} {et~al.}(2019){Klement}, {Carciofi}, {Rivinius}, {Ignace}, {Matthews}, {Torstensson}, {Gies}, {Vieira}, {Richardson}, {Domiciano de Souza}, {Bjorkman}, {Hallinan}, {Faes}, {Mota}, {Gullingsrud}, {de Breuck}, {Kervella}, {Cur{\'e}}, \& {Gunawan}}]{2019ApJ...885..147K}
{Klement}, R., {Carciofi}, A.~C., {Rivinius}, T., {et~al.} 2019, \bibinfo{title}{{Prevalence of SED Turndown among Classical Be Stars: Are All Be Stars Close Binaries?},} \apj, 885, 147, \dodoi{10.3847/1538-4357/ab48e7}

\bibitem[{R. {Klement} {et~al.}(2024){Klement}, {Rivinius}, {Gies}, {Baade}, {M{\'e}rand}, {Monnier}, {Schaefer}, {Lanthermann}, {Anugu}, {Kraus}, \& {Gardner}}]{2024ApJ...962...70K}
{Klement}, R., {Rivinius}, T., {Gies}, D.~R., {et~al.} 2024, \bibinfo{title}{{The CHARA Array Interferometric Program on the Multiplicity of Classical Be Stars: New Detections and Orbits of Stripped Subdwarf Companions},} \apj, 962, 70, \dodoi{10.3847/1538-4357/ad13ec}

\bibitem[{R. {Klement} {et~al.}(2025){Klement}, {Rivinius}, {Baade}, {M{\'e}rand}, {Bodensteiner}, {Frost}, {Sana}, {Shenar}, {Gies}, \& {Hadrava}}]{2025A&A...694A.208K}
{Klement}, R., {Rivinius}, T., {Baade}, D., {et~al.} 2025, \bibinfo{title}{{VLTI/GRAVITY enables the determination of the first dynamical masses of a classical Be + stripped and bloated pre-subdwarf binary},} \aap, 694, A208, \dodoi{10.1051/0004-6361/202453248}

\bibitem[{M. Livio \& E. Villaver(2009)Livio \& Villaver}]{Livio_Villaver_2009}
Livio, M., \& Villaver, E., eds. 2009, Massive Stars: From Pop III and GRBs to the Milky Way, Space Telescope Science Institute Symposium Series (Cambridge University Press)

\bibitem[{A. {Lobel} \& R. {Blomme}(2008){Lobel} \& {Blomme}}]{2008ApJ...678..408L}
{Lobel}, A., \& {Blomme}, R. 2008, \bibinfo{title}{{Modeling Ultraviolet Wind Line Variability in Massive Hot Stars},} \apj, 678, 408, \dodoi{10.1086/529129}

\bibitem[{A. {Maeder}(1996){Maeder}}]{1996LIACo..33...39M}
{Maeder}, A. 1996, in Liege International Astrophysical Colloquia, Vol.~33, Liege International Astrophysical Colloquia, ed. J.~M. {Vreux}, A.~{Detal}, D.~{Fraipont-Caro}, E.~{Gosset}, \& G.~{Rauw}, 39

\bibitem[{L. {Mahy} {et~al.}(2022){Mahy}, {Lanthermann}, {Hutsem{\'e}kers}, {Kluska}, {Lobel}, {Manick}, {Miszalski}, {Reggiani}, {Sana}, \& {Gosset}}]{2022A&A...657A...4M}
{Mahy}, L., {Lanthermann}, C., {Hutsem{\'e}kers}, D., {et~al.} 2022, \bibinfo{title}{{Multiplicity of Galactic luminous blue variable stars},} \aap, 657, A4, \dodoi{10.1051/0004-6361/202040062}

\bibitem[{P. {Marchant} \& J. {Bodensteiner}(2024){Marchant} \& {Bodensteiner}}]{2024ARA&A..62...21M}
{Marchant}, P., \& {Bodensteiner}, J. 2024, \bibinfo{title}{{The Evolution of Massive Binary Stars},} \araa, 62, 21, \dodoi{10.1146/annurev-astro-052722-105936}

\bibitem[{M. {Moe} \& R. {Di Stefano}(2017){Moe} \& {Di Stefano}}]{Moe2017}
{Moe}, M., \& {Di Stefano}, R. 2017, \bibinfo{title}{{Mind Your Ps and Qs: The Interrelation between Period (P) and Mass-ratio (Q) Distributions of Binary Stars},} \apjs, 230, 15, \dodoi{10.3847/1538-4365/aa6fb6}

\bibitem[{D.~J. {Mullan}(1984){Mullan}}]{1984ApJ...283..303M}
{Mullan}, D.~J. 1984, \bibinfo{title}{{Corotating interaction regions in stellar winds},} \apj, 283, 303, \dodoi{10.1086/162307}

\bibitem[{V. Muntean {et~al.}(2009)Muntean, Moffat, Chené, \& De~La~Chevrotière}]{Muntean2009}
Muntean, V., Moffat, A. F.~J., Chené, A.~N., \& De~La~Chevrotière, A. 2009, \bibinfo{title}{The Galactic hybrid Wolf–Rayet WN 7o/CE + O7V((f)) binary system WR 145,} Monthly Notices of the Royal Astronomical Society, 399, 1977, \dodoi{10.1111/j.1365-2966.2009.15288.x}

\bibitem[{V.~S. {Niemela} {et~al.}(1995){Niemela}, {Cabanne}, \& {Bassino}}]{Niemela1995}
{Niemela}, V.~S., {Cabanne}, M.~L., \& {Bassino}, L.~P. 1995, \bibinfo{title}{{The O5-7+WN binary system HDE 320102.},} \rmxaa, 31, 45

\bibitem[{V.~S. {Niemela} {et~al.}(1996){Niemela}, {Rovero}, \& {Cerruti}}]{Niemela1996}
{Niemela}, V.~S., {Rovero}, A.~C., \& {Cerruti}, M.~A. 1996, in Revista Mexicana de Astronomia y Astrofisica Conference Series, Vol.~5, Revista Mexicana de Astronomia y Astrofisica Conference Series, ed. V.~{Niemela}, N.~{Morrell}, P.~{Pismis}, \& S.~{Torres-Peimbert}, 126--128

\bibitem[{R.~K. {Prinja} {et~al.}(1990){Prinja}, {Barlow}, \& {Howarth}}]{1990ApJ...361..607P}
{Prinja}, R.~K., {Barlow}, M.~J., \& {Howarth}, I.~D. 1990, \bibinfo{title}{{Terminal Velocities for a Large Sample of O Stars, B Supergiants, and Wolf-Rayet Stars},} \apj, 361, 607, \dodoi{10.1086/169224}

\bibitem[{R.~K. {Prinja} \& I.~D. {Howarth}(1986){Prinja} \& {Howarth}}]{1986ApJS...61..357P}
{Prinja}, R.~K., \& {Howarth}, I.~D. 1986, \bibinfo{title}{{Narrow Absorption Components and Variability in Ultraviolet P Cygni Profiles of Early-Type Stars},} \apjs, 61, 357, \dodoi{10.1086/191117}

\bibitem[{T. {Ramiaramanantsoa} {et~al.}(2018){Ramiaramanantsoa}, {Moffat}, {Harmon}, {Ignace}, {St-Louis}, {Vanbeveren}, {Shenar}, {Pablo}, {Richardson}, {Howarth}, {Stevens}, {Piaulet}, {St-Jean}, {Eversberg}, {Pigulski}, {Popowicz}, {Kuschnig}, {Zoc{\l}o{\'n}ska}, {Buysschaert}, {Handler}, {Weiss}, {Wade}, {Rucinski}, {Zwintz}, {Luckas}, {Heathcote}, {Cacella}, {Powles}, {Locke}, {Bohlsen}, {Chen{\'e}}, {Miszalski}, {Waldron}, {Kotze}, {Kotze}, \& {B{\"o}hm}}]{2018MNRAS.473.5532R}
{Ramiaramanantsoa}, T., {Moffat}, A. F.~J., {Harmon}, R., {et~al.} 2018, \bibinfo{title}{{BRITE-Constellation high-precision time-dependent photometry of the early O-type supergiant {\ensuremath{\zeta}} Puppis unveils the photospheric drivers of its small- and large-scale wind structures},} \mnras, 473, 5532, \dodoi{10.1093/mnras/stx2671}

\bibitem[{M. {Renzo} {et~al.}(2019){Renzo}, {Zapartas}, {de Mink}, {G{\"o}tberg}, {Justham}, {Farmer}, {Izzard}, {Toonen}, \& {Sana}}]{2019A&A...624A..66R}
{Renzo}, M., {Zapartas}, E., {de Mink}, S.~E., {et~al.} 2019, \bibinfo{title}{{Massive runaway and walkaway stars. A study of the kinematical imprints of the physical processes governing the evolution and explosion of their binary progenitors},} \aap, 624, A66, \dodoi{10.1051/0004-6361/201833297}

\bibitem[{N.~D. {Richardson} {et~al.}(2024){Richardson}, {Schaefer}, {Eldridge}, {Spejcher}, {Holdsworth}, {Lau}, {Monnier}, {Moffat}, {Weigelt}, {Williams}, {Kraus}, {Le Bouquin}, {Anugu}, {Chhabra}, {Codron}, {Ennis}, {Gardner}, {Gutierrez}, {Ibrahim}, {Labdon}, {Lanthermann}, \& {Setterholm}}]{Richardson2024}
{Richardson}, N.~D., {Schaefer}, G.~H., {Eldridge}, J.~J., {et~al.} 2024, \bibinfo{title}{{Visual Orbits of Wolf{\textendash}Rayet Stars. I. The Orbit of the Dust-producing Wolf{\textendash}Rayet Binary WR 137 Measured with the CHARA Array},} \apj, 977, 78, \dodoi{10.3847/1538-4357/ad8d5c}

\bibitem[{T. {Rivinius} {et~al.}(2013){Rivinius}, {Carciofi}, \& {Martayan}}]{2013A&ARv..21...69R}
{Rivinius}, T., {Carciofi}, A.~C., \& {Martayan}, C. 2013, \bibinfo{title}{{Classical Be stars. Rapidly rotating B stars with viscous Keplerian decretion disks},} \aapr, 21, 69, \dodoi{10.1007/s00159-013-0069-0}

\bibitem[{C. {Rodrigo} \& E. {Solano}(2020){Rodrigo} \& {Solano}}]{SVO-2}
{Rodrigo}, C., \& {Solano}, E. 2020, in XIV.0 Scientific Meeting (virtual) of the Spanish Astronomical Society, 182

\bibitem[{C. {Rodrigo} {et~al.}(2012){Rodrigo}, {Solano}, \& {Bayo}}]{SVO-1}
{Rodrigo}, C., {Solano}, E., \& {Bayo}, A. 2012, \bibinfo{title}{{SVO Filter Profile Service Version 1.0},}, IVOA Working Draft 15 October 2012 \dodoi{10.5479/ADS/bib/2012ivoa.rept.1015R}

\bibitem[{C. {Rodrigo} {et~al.}(2024){Rodrigo}, {Cruz}, {Aguilar}, {Aller}, {Solano}, {G{\'a}lvez-Ortiz}, {Jim{\'e}nez-Esteban}, {Mas-Buitrago}, {Bayo}, {Cort{\'e}s-Contreras}, {Murillo-Ojeda}, {Bonoli}, {Cenarro}, {Dupke}, {L{\'o}pez-Sanjuan}, {Mar{\'\i}n-Franch}, {de Oliveira}, {Moles}, {Taylor}, {Varela}, \& {Rami{\'o}}}]{SVO-3}
{Rodrigo}, C., {Cruz}, P., {Aguilar}, J.~F., {et~al.} 2024, \bibinfo{title}{{Photometric segregation of dwarf and giant FGK stars using the SVO Filter Profile Service and photometric tools},} \aap, 689, A93, \dodoi{10.1051/0004-6361/202449998}

\bibitem[{H. {Sana} {et~al.}(2012){Sana}, {de Mink}, {de Koter}, {Langer}, {Evans}, {Gieles}, {Gosset}, {Izzard}, {Le Bouquin}, \& {Schneider}}]{Sana2012}
{Sana}, H., {de Mink}, S.~E., {de Koter}, A., {et~al.} 2012, \bibinfo{title}{{Binary Interaction Dominates the Evolution of Massive Stars},} Science, 337, 444, \dodoi{10.1126/science.1223344}

\bibitem[{H. {Sana} {et~al.}(2014){Sana}, {Le Bouquin}, {Lacour}, {Berger}, {Duvert}, {Gauchet}, {Norris}, {Olofsson}, {Pickel}, {Zins}, {Absil}, {de Koter}, {Kratter}, {Schnurr}, \& {Zinnecker}}]{Sana2014}
{Sana}, H., {Le Bouquin}, J.~B., {Lacour}, S., {et~al.} 2014, \bibinfo{title}{{Southern Massive Stars at High Angular Resolution: Observational Campaign and Companion Detection},} \apjs, 215, 15, \dodoi{10.1088/0067-0049/215/1/15}

\bibitem[{A.~A.~C. {Sander} {et~al.}(2019){Sander}, {Hamann}, {Todt}, {Hainich}, {Shenar}, {Ramachandran}, \& {Oskinova}}]{2019A&A...621A..92S}
{Sander}, A.~A.~C., {Hamann}, W.~R., {Todt}, H., {et~al.} 2019, \bibinfo{title}{{The Galactic WC and WO stars. The impact of revised distances from Gaia DR2 and their role as massive black hole progenitors},} \aap, 621, A92, \dodoi{10.1051/0004-6361/201833712}

\bibitem[{R.~E. {Schulte-Ladbeck} {et~al.}(1994){Schulte-Ladbeck}, {Clayton}, {Hillier}, {Harries}, \& {Howarth}}]{1994ApJ...429..846S}
{Schulte-Ladbeck}, R.~E., {Clayton}, G.~C., {Hillier}, D.~J., {Harries}, T.~J., \& {Howarth}, I.~D. 1994, \bibinfo{title}{{The Axisymmetric Stellar Wind of AG Carinae},} \apj, 429, 846, \dodoi{10.1086/174369}

\bibitem[{R.~E. {Schulte-Ladbeck} {et~al.}(1993){Schulte-Ladbeck}, {Leitherer}, {Clayton}, {Robert}, {Meade}, {Drissen}, {Nota}, \& {Schmutz}}]{1993ApJ...407..723S}
{Schulte-Ladbeck}, R.~E., {Leitherer}, C., {Clayton}, G.~C., {et~al.} 1993, \bibinfo{title}{{The Asymmetric Wind of R127},} \apj, 407, 723, \dodoi{10.1086/172553}

\bibitem[{I. {Shaposhnikov} {et~al.}(2023){Shaposhnikov}, {Cherepashchuk}, {Dodin}, \& {Postnov}}]{2023MNRAS.523.1524S}
{Shaposhnikov}, I., {Cherepashchuk}, A., {Dodin}, A., \& {Postnov}, K. 2023, \bibinfo{title}{{Spectroscopic searches for evolutionary orbital period changes in WR + OB binaries: the case of CQ Cep and CX Cep},} \mnras, 523, 1524, \dodoi{10.1093/mnras/stad1491}

\bibitem[{M.~M. {Shara} {et~al.}(2022){Shara}, {Howell}, {Furlan}, {Gnilka}, {Moffat}, {Scott}, \& {Zurek}}]{Shara2022}
{Shara}, M.~M., {Howell}, S.~B., {Furlan}, E., {et~al.} 2022, \bibinfo{title}{{A speckle-imaging search for close and very faint companions to the nearest and brightest Wolf-Rayet stars},} \mnras, 509, 2897, \dodoi{10.1093/mnras/stab2666}

\bibitem[{L.~F. {Smith} {et~al.}(1996){Smith}, {Shara}, \& {Moffat}}]{1996MNRAS.281..163S}
{Smith}, L.~F., {Shara}, M.~M., \& {Moffat}, A. F.~J. 1996, \bibinfo{title}{{A three-dimensional classification for WN stars},} \mnras, 281, 163, \dodoi{10.1093/mnras/281.1.163}

\bibitem[{L.~J. {Smith} {et~al.}(1994){Smith}, {Crowther}, \& {Prinja}}]{Smith94}
{Smith}, L.~J., {Crowther}, P.~A., \& {Prinja}, R.~K. 1994, \bibinfo{title}{{A study of the luminous blue variable candidate He 3-519 and its surrounding nebula.},} \aap, 281, 833

\bibitem[{N. {Smith}(2017){Smith}}]{Smith2017_rev}
{Smith}, N. 2017, \bibinfo{title}{{Luminous blue variables and the fates of very massive stars},} Philosophical Transactions of the Royal Society of London Series A, 375, 20160268, \dodoi{10.1098/rsta.2016.0268}

\bibitem[{N. {Smith} {et~al.}(2019){Smith}, {Aghakhanloo}, {Murphy}, {Drout}, {Stassun}, \& {Groh}}]{2019MNRAS.488.1760S}
{Smith}, N., {Aghakhanloo}, M., {Murphy}, J.~W., {et~al.} 2019, \bibinfo{title}{{On the Gaia DR2 distances for Galactic luminous blue variables},} \mnras, 488, 1760, \dodoi{10.1093/mnras/stz1712}

\bibitem[{N. {Smith} {et~al.}(2011){Smith}, {Li}, {Silverman}, {Ganeshalingam}, \& {Filippenko}}]{2011MNRAS.415..773S}
{Smith}, N., {Li}, W., {Silverman}, J.~M., {Ganeshalingam}, M., \& {Filippenko}, A.~V. 2011, \bibinfo{title}{{Luminous blue variable eruptions and related transients: diversity of progenitors and outburst properties},} \mnras, 415, 773, \dodoi{10.1111/j.1365-2966.2011.18763.x}

\bibitem[{N. {Smith} \& R. {Tombleson}(2015){Smith} \& {Tombleson}}]{2015MNRAS.447..598S}
{Smith}, N., \& {Tombleson}, R. 2015, \bibinfo{title}{{Luminous blue variables are antisocial: their isolation implies that they are kicked mass gainers in binary evolution},} \mnras, 447, 598, \dodoi{10.1093/mnras/stu2430}

\bibitem[{N. {St-Louis} {et~al.}(2009){St-Louis}, {Chen{\'e}}, {Schnurr}, \& {Nicol}}]{2009ApJ...698.1951S}
{St-Louis}, N., {Chen{\'e}}, A.~N., {Schnurr}, O., \& {Nicol}, M.~H. 2009, \bibinfo{title}{{A Systematic Search for Corotating Interaction Regions in Apparently Single Galactic Wolf-Rayet Stars. I. Characterizing the Variability},} \apj, 698, 1951, \dodoi{10.1088/0004-637X/698/2/1951}

\bibitem[{N. {St-Louis} {et~al.}(2018){St-Louis}, {Tremblay}, \& {Ignace}}]{2018MNRAS.474.1886S}
{St-Louis}, N., {Tremblay}, P., \& {Ignace}, R. 2018, \bibinfo{title}{{Polarization light curve modelling of corotating interaction regions in the wind of the Wolf-Rayet star WR 6},} \mnras, 474, 1886, \dodoi{10.1093/mnras/stx2813}

\bibitem[{N. {Steenken} {et~al.}(2025){Steenken}, {Ignace}, {St-Louis}, \& {Lenoir-Craig}}]{2025RNAAS...9...36S}
{Steenken}, N., {Ignace}, R., {St-Louis}, N., \& {Lenoir-Craig}, G. 2025, \bibinfo{title}{{Optical Polarimetry of WR 1 Consistent with a CIR and 17 days Rotation},} Research Notes of the American Astronomical Society, 9, 36, \dodoi{10.3847/2515-5172/adb4ea}

\bibitem[{I.~R. {Stevens} {et~al.}(1992){Stevens}, {Blondin}, \& {Pollock}}]{1992ApJ...386..265S}
{Stevens}, I.~R., {Blondin}, J.~M., \& {Pollock}, A.~M.~T. 1992, \bibinfo{title}{{Colliding Winds from Early-Type Stars in Binary Systems},} \apj, 386, 265, \dodoi{10.1086/171013}

\bibitem[{J.~A. {Toal{\'a}} {et~al.}(2015){Toal{\'a}}, {Guerrero}, {Ramos-Larios}, \& {Guzm{\'a}n}}]{2015A&A...578A..66T}
{Toal{\'a}}, J.~A., {Guerrero}, M.~A., {Ramos-Larios}, G., \& {Guzm{\'a}n}, V. 2015, \bibinfo{title}{{WISE morphological study of Wolf-Rayet nebulae},} \aap, 578, A66, \dodoi{10.1051/0004-6361/201525706}

\bibitem[{S. {van der Walt} {et~al.}(2011){van der Walt}, {Colbert}, \& {Varoquaux}}]{numpy}
{van der Walt}, S., {Colbert}, S.~C., \& {Varoquaux}, G. 2011, \bibinfo{title}{{The NumPy Array: A Structure for Efficient Numerical Computation},} Computing in Science and Engineering, 13, 22, \dodoi{10.1109/MCSE.2011.37}

\bibitem[{A.~M. {van Genderen}(2001){van Genderen}}]{2001A&A...366..508V}
{van Genderen}, A.~M. 2001, \bibinfo{title}{{S Doradus variables in the Galaxy and the Magellanic Clouds},} \aap, 366, 508, \dodoi{10.1051/0004-6361:20000022}

\bibitem[{A. {Villar-Sbaffi} {et~al.}(2005){Villar-Sbaffi}, {St-Louis}, {Moffat}, \& {Piirola}}]{2005ApJ...623.1092V}
{Villar-Sbaffi}, A., {St-Louis}, N., {Moffat}, A. F.~J., \& {Piirola}, V. 2005, \bibinfo{title}{{First Ever Polarimetric Detection of a Wind-Wind Interaction Region and a Misaligned Flattening of the Wind in the Wolf-Rayet Binary CQ Cephei},} \apj, 623, 1092, \dodoi{10.1086/428830}

\bibitem[{J.~S. {Vink}(2018){Vink}}]{Vink2018}
{Vink}, J.~S. 2018, \bibinfo{title}{{Very massive stars: a metallicity-dependent upper-mass limit, slow winds, and the self-enrichment of globular clusters},} \aap, 615, A119, \dodoi{10.1051/0004-6361/201832773}

\bibitem[{J.~S. {Vink} {et~al.}(2011){Vink}, {Gr{\"a}fener}, \& {Harries}}]{2011A&A...536L..10V}
{Vink}, J.~S., {Gr{\"a}fener}, G., \& {Harries}, T.~J. 2011, \bibinfo{title}{{In pursuit of gamma-ray burst progenitors: the identification of a sub-population of rotating Wolf-Rayet stars},} \aap, 536, L10, \dodoi{10.1051/0004-6361/201118197}

\bibitem[{K. {Weis} \& D.~J. {Bomans}(2020){Weis} \& {Bomans}}]{2020Galax...8...20W}
{Weis}, K., \& {Bomans}, D.~J. 2020, \bibinfo{title}{{Luminous Blue Variables},} Galaxies, 8, 20, \dodoi{10.3390/galaxies8010020}

\bibitem[{T. Williams {et~al.}(2023)Williams, Kelley, Lang, Kotz, Campbell, Elber, Woo, \& {many others}}]{gnuplot}
Williams, T., Kelley, C., Lang, R., {et~al.} 2023, \bibinfo{title}{gnuplot 5.4,}, \url{https://sourceforge.net/projects/gnuplot/files/gnuplot/}

\bibitem[{J.~P. {Wisniewski} {et~al.}(2006){Wisniewski}, {Babler}, {Bjorkman}, {Kurchakov}, {Meade}, \& {Miroshnichenko}}]{2006PASP..118..820W}
{Wisniewski}, J.~P., {Babler}, B.~L., {Bjorkman}, K.~S., {et~al.} 2006, \bibinfo{title}{{The Asymmetrical Wind of the Candidate Luminous Blue Variable MWC 314},} \pasp, 118, 820, \dodoi{10.1086/506182}

\end{thebibliography}

\end{document}